\documentclass[reprint,amsmath,amssymb,superscriptaddress,nolongbibliography,aps,prl,nofootinbib]{revtex4-2}

\usepackage{xr}
\usepackage[utf8]{inputenc}
\usepackage{amsmath}
\usepackage{amsmath,amssymb}
\usepackage{siunitx}
\usepackage[dvipsnames]{xcolor}
\usepackage{color,soul}
\setstcolor{red}
\sethlcolor{yellow}
\usepackage{notes2bib}
\usepackage[colorlinks=true]{hyperref}
\hypersetup{
    allcolors = {blue},
}

\newenvironment{eqs}%
{\begin{equation} \begin{aligned}}%
{\end{aligned} \end{equation} }
\newcommand{\beal}{\begin{eqs}}
\newcommand{\eal}{\end{eqs}}

\usepackage{graphicx}
\usepackage{dcolumn}
\usepackage{bm}
\usepackage{booktabs}
\usepackage{framed,enumitem}

\begin{document}
\preprint{APS/123-QED}

\title{Universal Moir\'e Buckling of {Freestanding} 2D Bilayers}

\author{Jin Wang}
    \affiliation{International School for Advanced Studies (SISSA), Via Bonomea 265, 34136 Trieste, Italy}
    \affiliation{International Centre for Theoretical Physics (ICTP), Strada Costiera 11,34151 Trieste,Italy}
\author{Erio Tosatti}
\email{tosatti@sissa.it}
    \affiliation{International School for Advanced Studies (SISSA), Via Bonomea 265, 34136 Trieste, Italy}
    \affiliation{International Centre for Theoretical Physics (ICTP), Strada Costiera 11,34151 Trieste,Italy}


\begin{abstract}
The physics of membranes, a classic subject, acquires new momentum from two-dimensional (2D) materials multilayers. This work reports the suprising results emerged during  a theoretical study of equilibrium geometry of bilayers as freestanding membranes. 
While ordinary membranes are prone to buckle around compressive impurities, we predict that all 2D material freestanding bilayers universally undergo, even if impurity-free, a spontaneous out-of-plane buckling.
The moir\'e network nodes here play the role of {\it internal} impurities, the dislocations that join them giving rise to a stress pattern, purely shear in homo-bilayers and mixed compressive/shear in hetero-bilayers. That intrinsic stress is, theory and simulations show, generally capable to cause all freestanding 2D bilayers to undergo distortive bucklings with large amplitudes and a rich predicted phase transition scenario.
Realistic simulations predict quantitative parameters expected for {these phenomena as expected in} hetero-bilayers such as graphene/hBN, $\rm{WS_2/WSe_2}$ hetero-bilayers, and for twisted homo-bilayers such as graphene, hBN, $\rm{MoS_2}$.
Buckling then entails a variety of predicted consequences. Mechanically, a critical drop of bending stiffness is expected at all buckling transitions. Thermally, the average buckling corrugation decreases with temperature, with buckling-unbuckling phase transitions expected in some cases, 
and the buckled state often persisting even above room temperature.
Buckling will be suppressed by deposition on hard attractive substrates, and survive in reduced form on soft ones.
Frictional, electronic and other associated phenomena are also highlighted. The universality and richness of these predicted phenomena strongly encourages an experimental search, which is possible but still missing.

\end{abstract}

\maketitle

The surge of interest in the physics of 2D materials \cite{Geim.natmater.2007,Geim.nature.2013,Neto.science.2016} also raises important questions about their structural response as membranes under mechanical perturbations \cite{Wang.prl.2019,Wang.jem.2023,Hou.nc.2024}.
Elegant theory work has described in literature the effect of embedded, extrinsic impurities creating a local compressive stress in a freestanding crystalline membrane, such as 2D monolayers \cite{Plummer.pre.2020,Plummer.prm.2022} but also spheres and other systems and geometries \cite{Nelson.pre.2003, Nelson.pre.2007}.
Above a critical compression magnitude exerted by an impurity, the membrane shape spontaneously changes from flat to an out-of-plane buckled state,
{a type of shape morphing \cite{Sharon.2004,Klein.science.2007,Efrati.jmps.2009} that has been widely observed \cite{Kim.science.2012,Wu.natcommu.2013,xu.prl.2020}.}
An array of impurities leads to a buckling superlattice, generally {found to develop} anti-ferro ordering between neighboring impurity sites \cite{Plummer.pre.2020}.
Do {\it shear} perturbations also occur, and can they lead to buckling in 2D materials {and their stacks}?
May large amplitude bucklings occur spontaneously and intrinsically rather than extrinsically and artificially? We show here that both answers are universally affirmative in 2D twisted bilayers, where the moir\'e pattern acts as an {\it internal} perturbation, generally introducing shear stress and not just compression.
Homo-bilayers such as twisted graphene \cite{Bistritzer.PNAS.2011, Cao.nature.2018-2, Andrei.natmater.2020}, hBN \cite{Stern.science.2021}, and transition metal dichalcogenides (TMD) \cite{Weston.natnano.2020},
as well as hetero-bilayers, like graphene/hBN \cite{Hunt.science.2013} and $\rm{WS_2/WSe_2}$ \cite{Li.natmater.2021}, achieved popularity for their exceptional
mechanical and electronic phenomena \cite{Yan.intermater.2024,Sun.ChemRev.2024}.
They come in a variety of twist angles $\theta$ and conditions ranging from fully encapsulated \cite{Cao.nature.2018-1, Cao.nature.2018-2}, to deposited \cite{Lisi.natphys.2021, Liao.natmater.2022}, to (more rarely) suspended \cite{Wang.prl.2017} and freestanding \cite{Butz.nature.2014, Woo.prb.2023}.
Here we demonstrate the universal structural buckling phenomena that will occur under appropriate twists in all freestanding 2D bilayers.\\

Mechanically, a twisted bilayer is a very unconventional membrane. For a majority of twist angles $\theta$ and in the absence of defects, steps or boundaries, the two layers are unpinned and can slide freely relative to one another \cite{Wang.rmp.2024}.
Each layer is homogeneous, but the bilayer as a whole possesses the intrinsic inhomogeneity provided by the moir\'e modulation, and that may lead to buckling.
The modest buckling magnitudes observed in deposited bilayers \cite{Li.natmater.2021}, where flatness is enforced by a substrate, have not attracted special attention besides revealing the moir\'e pattern.
Freestanding bilayers are instead free to behave as genuine membranes.
To date, no experimental information about their flatness or buckling seems available.
Addressing therefore the freestanding bilayers by theory and simulation, we encountered first one \cite{Wang.prb.2023}, then a whole class of intriguing phenomena which form the contents of this report. 
Where first of all the general physical picture is addressed, and subsequently specific quantitative predictions are made that should be relevant for future developments and experiments.
We do that in three steps. First, by introducing, at the mean-field theory level, buckling of a generic defect-free membrane under externally applied shear or compressive stress.
Second, demonstrating by simulation the buckling effect of a moir\'e-shaped non-uniform perturbation, either shear and compressive, on the so-called neutral plane of a hypothetical bilayer.
Third, predicting by full fledged realistic simulations the spontaneous low temperature buckling of some important 2D bilayers.\\

\begin{figure}[hb!]
\centering
\includegraphics[width=1\linewidth]{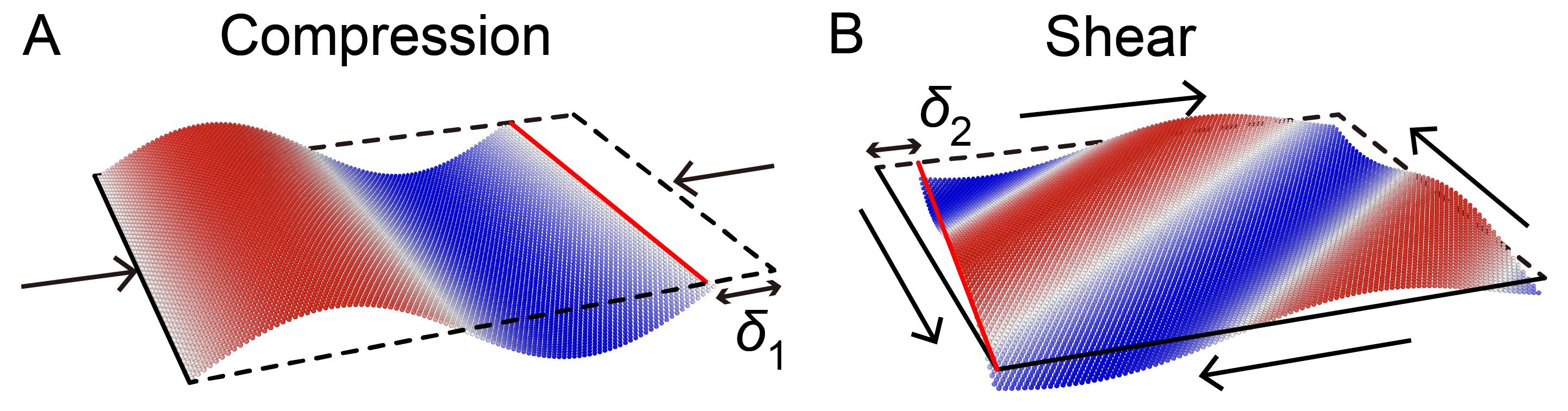}
\caption{Buckling of rectangular membranes due to (A) compression and (B) shear. In-plane compression and shear deformations are denoted by $\delta_1$ and $\delta_2$. Color shows the out-of-plane deformation. 
}
\label{fig:sketch}
\end{figure}

\section*{Uniformly stressed membrane}

The Föppl–von Kármán free energy $F$ of a generic $(x, y)$ 
defect-free membrane under an external perturbation \cite{Landau.elasticity}, is
$\frac{1}{2}\iint dx dy \big[ D (\nabla^2 h)^2 + \lambda (\varepsilon_{xx}+\varepsilon_{yy})^2 
+ 2 \mu(\varepsilon_{xx}^2+\varepsilon_{yy}^2+2\varepsilon_{xy}^2) -2d \sigma_{ij} \varepsilon_{ji}  \big ]$,
where the last term is the coupling to {the external} stress field $\sigma_{ij}$, $d$ is the effective thickness of the membrane, $h$ is the out-of-plane $z$-deformation, $D$ the bending stiffness, $\lambda$ and $\mu$ are the 2D lam\'e constants, and strain {is} $\varepsilon_{ij} = (1/2)[\partial u_i/\partial x_j +  \partial u_j/\partial x_i + (\partial h/\partial x_i)(\partial h/\partial x_j)]$,
with $u_x$ and $u_y$ the in-plane elastic deformations.
We consider the effect of uniform compression and shear perturbations caused  respectively by displacements $\delta_1$ and $\delta_2$, both $\ll L$, the overall membrane size, as 
in Fig.~\ref{fig:sketch}.
With a trial sinusoidal buckling
$h = h_0 \sin{(\frac{2\pi}{L}x)}$ (compressive case) and
$h = h_0 \sin{\left[ \frac{2\pi}{L} (x - y) \right]}$ (shear case), the minimized free energy for a square membrane of area $A=L^2$ is respectively
\begin{equation}
\begin{aligned}
   F_\mathrm{c} &= Y\delta_1^2/2 - \alpha Y \left( \delta_1 - \delta_\mathrm{1crit} \right)^2 \Theta{\left( \delta_1 - \delta_\mathrm{1crit} \right)}, \\
   F_\mathrm{s} &= \mu \delta_2^2/2 - \beta \mu \left( \delta_2 - \delta_\mathrm{2crit} \right)^2 \Theta {\left( \delta_2 - \delta_\mathrm{2crit} \right)},
\end{aligned}
\end{equation}
where $Y$ is Young's modulus, $\alpha=(1-\nu^2)/3$, $\beta=(1-\nu)/6$, $\nu$ is Poisson's ratio and $\Theta$ is Heaviside's step function.
At $\delta_1 = \delta_\mathrm{1crit} =4\pi^2 D/Y L$, there is a sharp second order buckling transition with magnitude
\begin{equation}
    h_0 = \mathrm{Re} \sqrt{ 2 \alpha \left( \delta_1 - \delta_\mathrm{1crit} \right) L/\pi^2}
\end{equation}
and similar for shear at $\delta_\mathrm{2crit} = 8\pi^2 D/\mu L$ (details in SI).
{A similar result was obtained in the earlier, more general  mean field treatment of Hanakata \textit{et al.} \cite{Hanakata.eml.2021} for thin ribbons under compression. A Landau theory of shear buckling could be constructed in a similar fashion.}
The critical Föppl–von Kármán number \cite{Plummer.pre.2020}
$\gamma_\mathrm{crit}$  equals here
$ \gamma_\mathrm{crit} = YL\delta_\mathrm{1crit}/D =4\pi^2$ for compressive buckling, and
$\mu L\delta_\mathrm{2crit}/D = 8\pi^2 $ for shear buckling.
This elementary uniform result also serves to clarify that shear stress and not only compressive stress will give rise to a buckling distortion above a critical threshold, lower for larger size $L$ and for weaker bending stiffness $D$.
Of course shear and compressive bucklings occur, as shown, 45$^{\circ}$ apart \cite{Cerda.prl.2003, Timoshenko}.
{However,} since typically $Y \approx 3\mu$, the shear-induced buckling requires a stronger perturbation compared to compression.

\begin{figure}[ht!]
\centering
\includegraphics[width=0.95\linewidth]{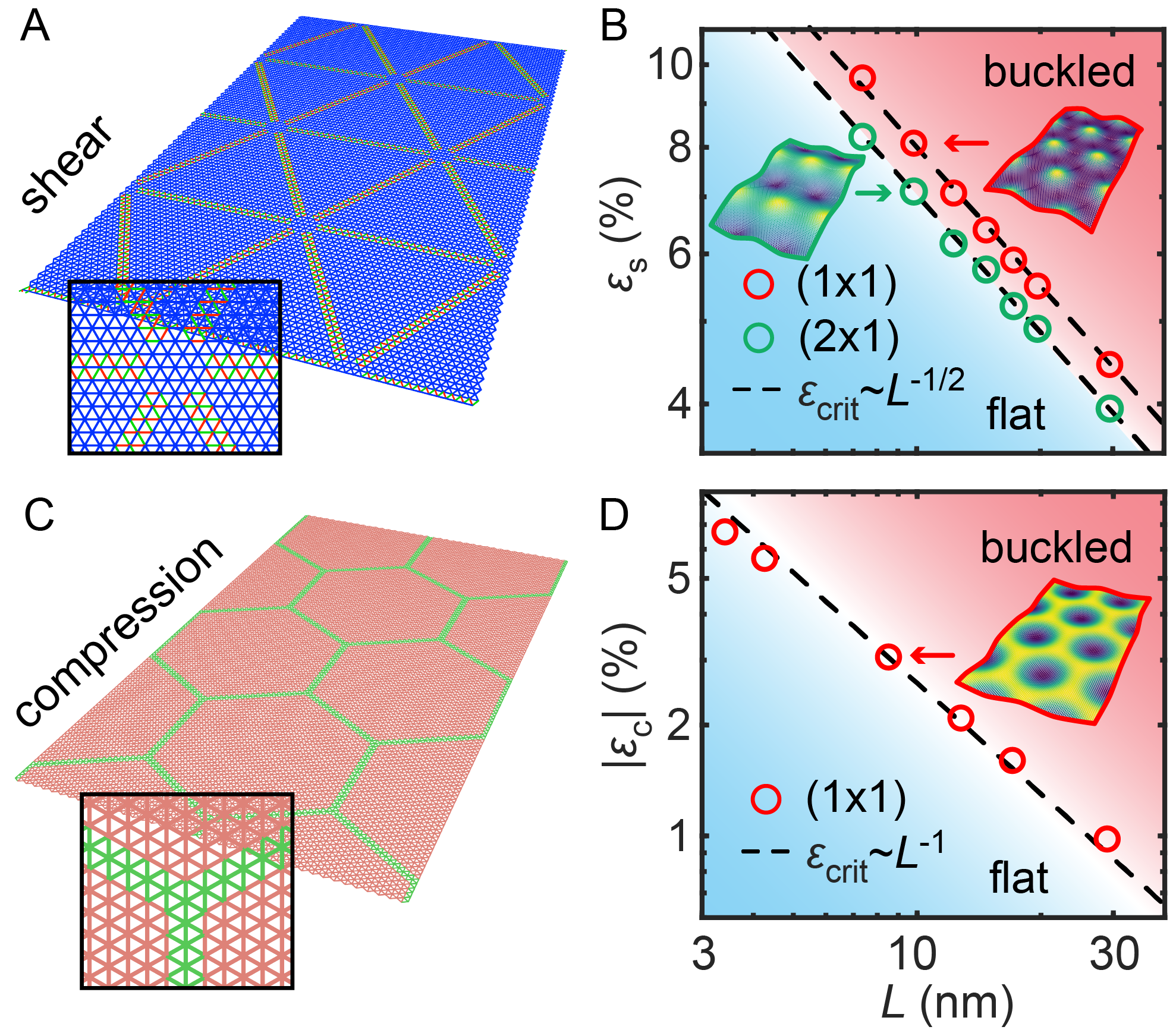}
\caption{
Neutral membrane models with added perturbing patterns (left panel) and resulting buckling phase diagrams (right panel) for shear (A and B) and compression (C and D) cases.
In (A, C), red, green and blue represent elongated, shortened and original bond length.
Optimized structures are shown in the inset of (B, D) with colors indicating out-of-plane corrugations.}
\label{fig:monolayer}
\end{figure}

\begin{figure*}[ht!]
\centering
\includegraphics[width=0.96\linewidth]{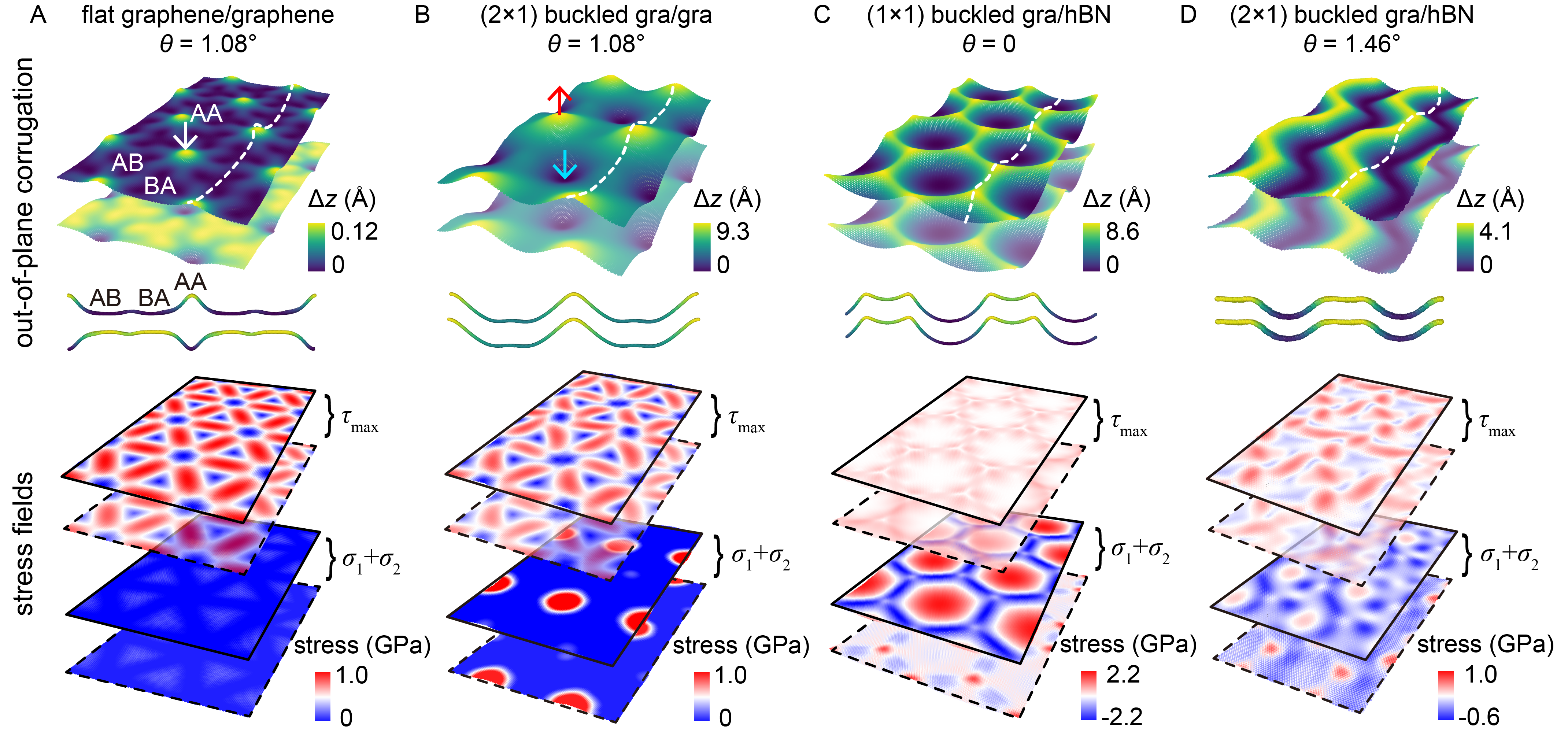}
\caption{Out-of-plane spontaneous $T=0$ corrugation patterns and stress fields from simulations of some real bilayers: 
{(A) homo graphene bilayer in its artificially constrained flat state, and (B) in its $(2\times 1)$ lowest energy buckled state; 
(C) hetero gra/hBN bilayer in $(1\times 1)$ buckled state at zero twist; 
(D) same as (C) in the $(2\times 1)$ buckled state at $\theta=1.46^\circ$.
Top, middle and lower panels: 2D map of corrugation $\Delta z=z-\min{\{z\}}$; $\Delta z$ along the white dashed line; shear and compressive stress fields. Designation of AA nodes and of AB, BA domains is illustrated in panel (A).
In the ``antiferro" $(2\times 1)$ buckled structure, the AA nodes buckle alternatively up and down, as indicated by the red/blue arrows in the top panel of (B). 
The compressive stress can be both positive and negative. The maximum shear stress $\tau_\mathrm{max}=(\sigma_2-\sigma_1)/2$ is instead always positive -- the two principal stresses being defined as $\sigma_2 \geq \sigma_1$. 
Results for $\rm{MoS_2/MoS_2}$ and for $\rm{WS_2/WSe_2}$ are qualitatively similar to those of panels (B) and (C) respectively.}}
\label{fig:stress}
\end{figure*}

\section*{Non-uniform stress -- {the} neutral membrane model}

The moir\'e-induced stress pattern is far from uniform in a bilayer, the system which we eventually wish to address.
Following Landau-Lifshitz \cite{Landau.elasticity}, the overall $z$-geometry of a bilayer or multilayer may be represented by that of an imaginary neutral membrane (Fig.~\ref{fig:monolayer}A).
Its possible buckling behaviour under the moir\'e-induced stress pattern can be qualitatively and numerically checked by a simple monolayer model, which we arbitrarily endow with a triangular crystalline structure.
{The model Hamiltonian is 
$\mathcal{H}=V_\mathrm{b}+V_\mathrm{a}+V_\mathrm{d}$, including two-body bond interactions, three-body angular interactions, and four-body dihedrals. Parameters are chosen to give this neutral membrane reasonable values of the bending stiffness $D$ and elastic moduli $Y$ and $\mu$  (details in Methods).}
Extending Plummer-Nelson's method \cite{Plummer.pre.2020}, we introduce stress in the neutral membrane through a moir\'e-like 2D network of intrinsic in-plane ``impurities'', created by modifying the equilibrium length $l_\mathrm{equ}$ of bonds within the network.
As we shall see, in the actual twisted bilayers \cite{Zhang.jmps.2018,Kazmierczak.NatMater.2021}, the moir\'e stress can be purely shear or mixed compressive/shear, depending on homo/hetero character, lattice mismatch $\delta$, and twist angle $\theta$.
To introduce local shear stress, the equilibrium length $l_\mathrm{equ}$ of two adjacent bonds at $120^{\circ}$ (Fig.~\ref{fig:monolayer}A) are artificially changed from
$a$ to $(1+\varepsilon_\mathrm{s}) a$ and $(1-\varepsilon_\mathrm{s}) a$ respectively.
To introduce compression (Fig.~\ref{fig:monolayer}C), bonds within the network are instead changed to $(1-\varepsilon_\mathrm{c})a$, compensated by a small increase in the rest to ensure unchanged overall initial size.

As a function of the perturbing strengths $\varepsilon$, structural optimization of the model neutral membrane (carried out while maintaining zero in-plane pressure) shows a spontaneous buckling above a critical $\varepsilon =  \varepsilon_\mathrm{crit}$, for pure shear as well as for compression (Fig.~\ref{fig:monolayer}B, D).
At $\varepsilon_\mathrm{crit}$ the bending energy cost $E_\mathrm{bend}\propto D$ matches the in-plane elastic energy gain per moir\'e,
$\Delta E_\mathrm{in} \propto L a\mu \varepsilon^2$ (shear) and
$ \propto L^2 Y \varepsilon^2$ (compression).
The reason for the different $L$ exponent is that the shear strain is narrowly localized at the network, while compressive strain spreads globally (details in SI).
This expected scaling with moir\'e size $L$ is confirmed by a numerical fit of simulation results, yielding
$\varepsilon_\mathrm{s,crit} \sim (D/\mu L a)^{1/2}$ and
$\varepsilon_\mathrm{c,crit} \sim (D/Y L^2)^{1/2}$ respectively (Fig.~\ref{fig:monolayer}B, D).
The strong non-uniformity of the perturbing stress is thus responsible for the 1/2 power instead of 1 in the uniform compressive case.

In twisted homo-bilayers, where the moir\'e-induced strain is strictly shear because the two layers are identical, $\varepsilon_s \to 0$ when $L \to \infty$, a limit attained for infinitesimal twist, $\theta \approx a/L \to 0$.
The perturbing shear stress, dependent on the moir\'e dislocation width $w$, increases as $\theta$ decreases, but saturates below $\theta \approx 1^{\circ}$ \cite{Wang.rmp.2024, Zhang.jmps.2018}. 
(The connection between $\epsilon$ and $w$, and the reason why $w$ saturates when $\theta \to $ 0, will be detailed in next Section.) 
As a consequence we anticipate that, remarkably, \textit{all freestanding homo-bilayers should spontaneously buckle at small twist angles} (at $T=0$).
The neutral membrane buckling geometry is antiferro, reflecting the up-down symmetry already noted in the homogeneous case.
We anticipate that, in that respect, hetero-bilayers  must  differ importantly from homo-bilayers. The lattice mismatch of the two partner layers implies an omnipresent moir\'e-induced compressive stress, plus, as in the homo case, a shear stress growing like $\theta$.
Because at $\theta=0$, $L \sim ab/|a-b|$ is finite ($a$ and $b$ being the lattice constant of two layers) \cite{Wang.rmp.2024}, the critical stress $\varepsilon_\mathrm{c,crit}$ may or may not be reached.
At $\theta$ = 0  the moir\'e induced stress is purely compressive, and buckling can only have ``ferro" symmetry, that is, $(1 \times 1)$ periodicity.
Finally, when $\theta \neq 0$ and a nonzero shear stress component enters, the resulting scenario is generally more complicated as we shall see.

\begin{figure*}[ht!]
\centering
\includegraphics[width=0.7\linewidth]{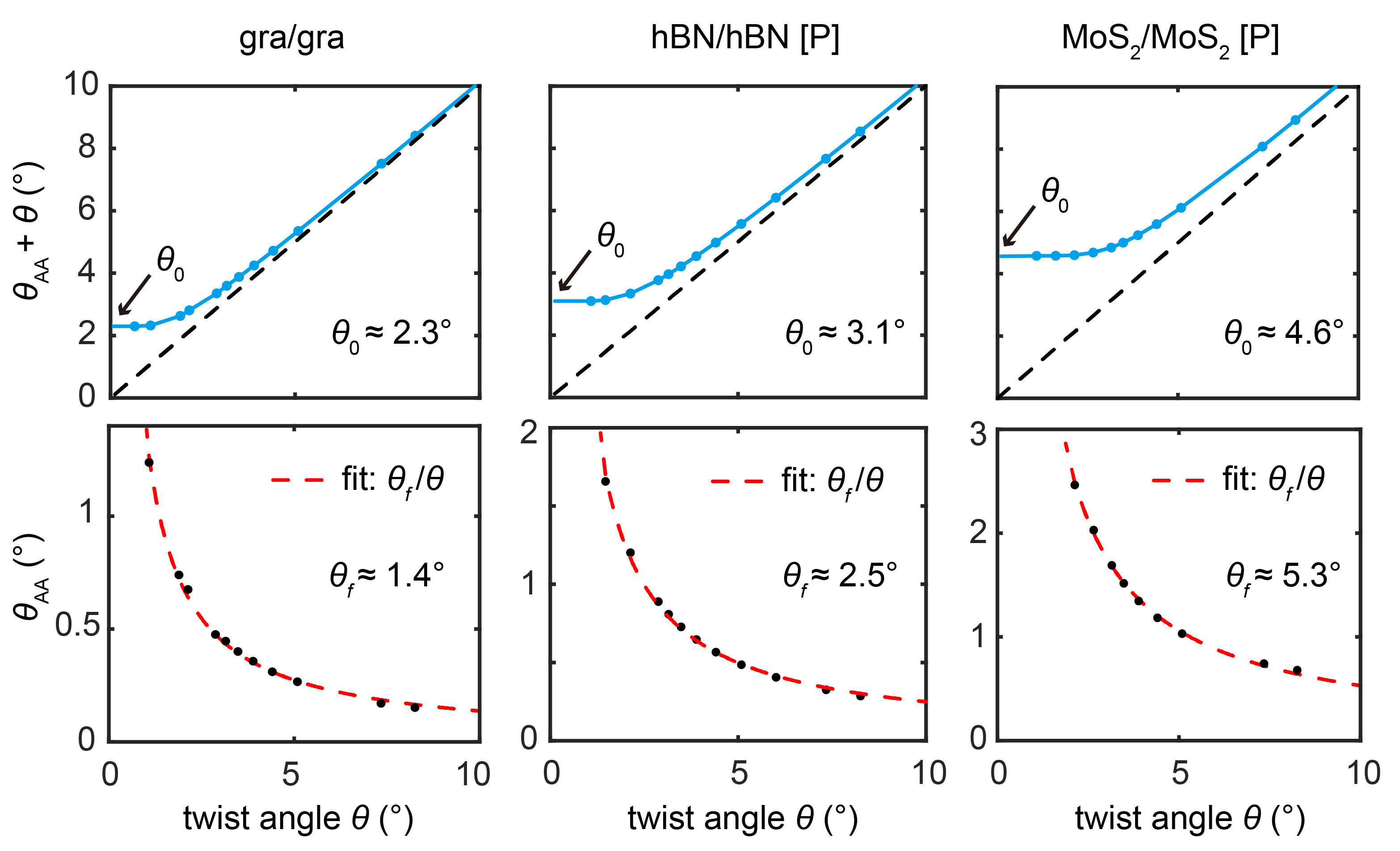}
\caption{Maximum local twist angle $\theta_\mathrm{max}=\theta_\mathrm{AA}+\theta$ and AA node ``twirl" $\theta_\mathrm{AA}$ for twisted homo-bilayers: graphene, parallel-stacked hBN and $\mathrm{MoS_2}$. For $\theta \gtrsim \theta_f/2$, $\theta_\mathrm{AA}$ are well represented by $\theta_f/\theta$.
In the $\theta \to 0$ limit, $\theta_\mathrm{max} \to \theta_\mathrm{AA} \to \theta_0$.
}
\label{fig:scaling}
\end{figure*}

\section*{Buckling in real 2D bilayers}

The suggestions provided by the above preparatory steps are now ready for verification by realistic simulations of true important bilayers. We explore the optimal equilibrium bilayer geometries in five cases with decreasing lattice mismatch $\delta$: 
two hetero-bilayers:
$\rm{WS_2/WSe_2}$ ($\delta \approx 5.12\%$) and graphene/hBN ($\delta \approx1.83\%$);
and three twisted homo-bilayers ($\delta=0$): graphene, hBN, and $\rm{MoS_2}$.
The lowest energy state is searched by simulation using well tested empirical force fields \cite{Ouyang.nanolett.2018,Jiang.JPCA.2023}.
The simulation protocol (details in Methods) allows all atomic coordinates to change freely, while enforcing zero overall in-plane stresses.
At small twist angles, all bilayers are found to buckle strongly and spontaneously, confirming expectations. Their overall shape (i.e., their neutral plane) bulges ``ferromagnetically'' up $(1 \times 1)$ in hetero-bilayers {(Fig.~\ref{fig:stress}C)}, but {alternate} ``anti-ferromagnetically'' up-down among neighboring moir\'e AA nodes in homo-bilayers {(Fig.~\ref{fig:stress}B)}. 
Different antiferro states {such as $(2 \times 1)$, $(\sqrt{3} \times \sqrt{3})$, $(2 \times 2)$} have similar $T$=0 energies, as will be exemplified by later for graphene/graphene at $\theta=1.89^{\circ}$.
While any one of these buckled state symmetries may prevail depending on hard-to-predict small perturbations, here and below we shall restrict our discussion to the $(2 \times 1)$ ``striped" state.
\footnote{
The ground state morphology makes it tempting to map, as was done in Ref.~\cite{Plummer.pre.2020}, the buckling physical situation to an Ising model on the triangular lattice.
In that Hamiltonian, $\mathcal{H} = -\Sigma_{ij} J_{ij} S_i S_j$, the first-neighbour $J_1 > 0$ accounts for prevalence of antiferro ordering, while $J_2$ and $J_3$ will determine in detail which one among several types of ordering should prevail at $T$=0 \cite{Tanaka.ptp.1976}.
Identifying ``spins" with AA nodes, and ignoring for the sake of qualitative understanding the different buckling magnitudes, differences between the $T$=0 energies, namely
$6J_1+ 6J_2+ 6J_3$ for the ferro $(1 \times 1)$ state;
$-2J_1-2J_2+6J_3$ and $-2J_1+2J_2-2J_3$ for the striped $(2 \times 1)$ and $(2\times 2)$ antiferro states;
$-2J_1+6J_2-2J_3$ for the ``ferrimagnetic" $(\sqrt{3} \times \sqrt{3})$ state \cite{Tanaka.ptp.1976}, permit the estimate of interaction values.
Using as an example the graphene/graphene with $\theta=1.89^{\circ}$, we get in units of meV/moir\'e $J_1$= 10.46, $J_2$= 3.56, $J_3$= 2.30 respectively in that case.
Curiously, we note that while thermal simulations of real bilayers never show  the fluctuating ``spin flips" expected of an Ising model, the magnitude of $2J_1$ and the real buckling-unbuckling $T_c \sim$ 300 K (Fig.~S3 of Ref.~\cite{Wang.prb.2023}) are not too far apart.}

The moir\'e perturbation of a homo-bilayer is strictly shear, with the shear stress concentrated at the moir\'e discommensuration network (Fig.~\ref{fig:stress}A). 
Geometrically, the perturbing {strain} magnitude is in turn connected to and controlled by the AA node local rotation (``twirling" \cite{Yoo.NatMater.2019,Kaliteevski.nanolett.2023}) angle $\theta_\mathrm{AA}$, in the form 
$\varepsilon_\mathrm{s} \sim a/8w \sim \theta_\mathrm{AA}$ \cite{Zhang.jmps.2018} -- the factor 8 in the denominator approximates the real graphene/graphene result. 
{The local rotation at AA nodes and local shear along the network occurs spontaneously after structural relaxation, whereby entire regions ``reconstructively" turn into commensurate AB and BA domains after optimization \cite{Yoo.NatMater.2019}. 
As a result, the overall twist concentrates at the AA nodes and the discommensuration network, leading to an increase in local rotation and shear.}
The connection between individual {AA} node twirling and shear {strain}
in the network between nodes follows from a straight mechanical analogy with tracks pulled by two rotating  wheels (Fig.~S3 in SI).

\begin{figure*}[ht!]
\centering
\includegraphics[width=0.8\linewidth]{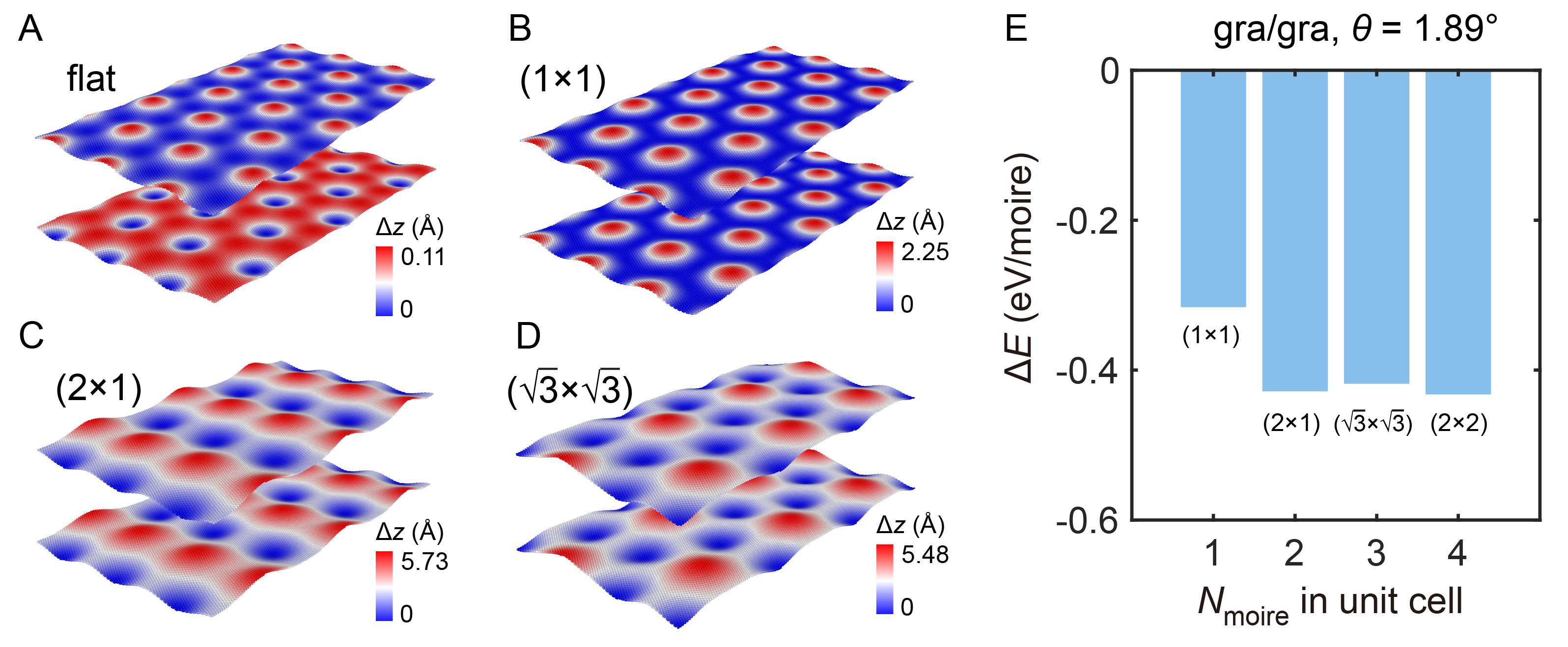}
\caption{Different optimized structures of homo-bilayer graphene with $\theta=1.89^{\circ}$:
A. supported-flat configuration; B. $(1\times 1)$ buckling, C. $(2\times 1)$ buckling, and D. $(\sqrt{3}\times \sqrt{3})$ buckling configuration. E. Energy difference (per-moire unit cell, $T$ = 0) for various locally stable configurations with respect to energy of the flat state.
}
\label{fig:energy}
\end{figure*}

\begin{figure}[ht!]
\centering
\includegraphics[width=0.8\linewidth]{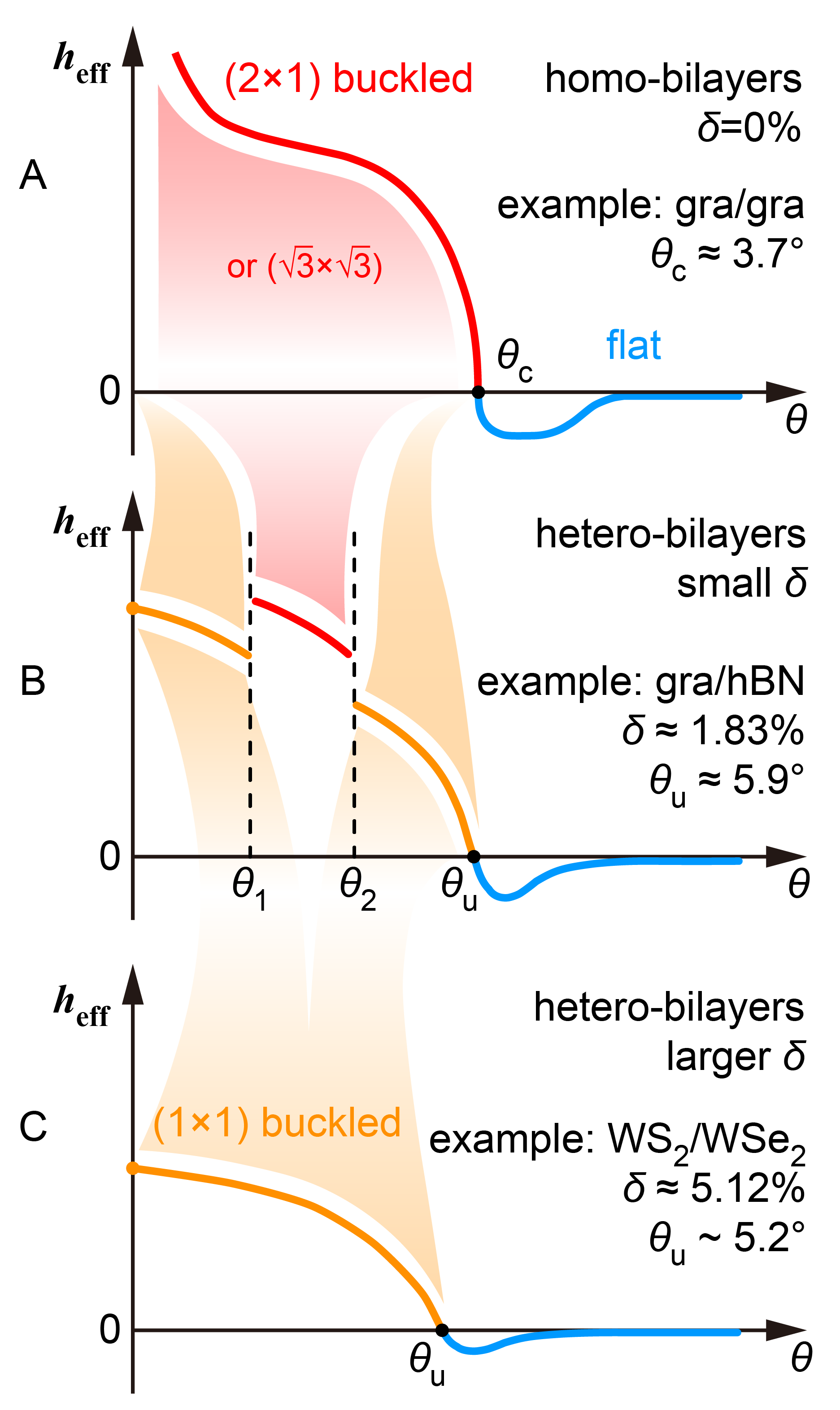}
\caption{
Universal  phase diagram (not to scale) of the $T$=0 buckling corrugation
$h_\mathrm{eff} = 2\sqrt{\max{|\Delta \vec{z}_\mathrm{up} \cdot \Delta \vec{z}_\mathrm{low}|}} \mathrm{sign}(\left <  \Delta \vec{z}_\mathrm{up} \cdot \Delta \vec{z}_\mathrm{low} \right >)$ of 2D freestanding bilayers as a function of twist angle $\theta$, where $\Delta \vec{z} = \vec{z} - \left<\vec{z} \right>$.
Negative values correspond to flat, unbuckled states.
{A. }For homo-bilayers, the twist angle of the buckled-unbuckled critical transition is denoted as $\theta_c$. 
{B-C. }For hetero-bilayers, $\theta_u$ is the buckled-unbuckled crossover twist angle. Only for small mismatch $\delta$ {(B)} there can be a $(2 \times 1)$ intruding phase.}
\label{fig:master}
\end{figure}

Negligible at large twist angle $\theta$, $\varepsilon_\mathrm{s}$ grows while $w$ drops while $\theta$ decreases, down to a crossover $\theta \approx \theta_f$ \cite{Wang.rmp.2024}.
Around and below $\theta_f$ the moir\'e distortion changes from weak and sinusoidal to strong and concentrated onto a network of narrow discommensurations -- an evolution sometimes called ``moir\'e reconstruction" \cite{Yoo.NatMater.2019}.
Below $\theta_f$, all quantities, $\theta_\mathrm{AA}$, $\varepsilon_\mathrm{s}$, and correspondingly $w$ eventually saturate to a finite value. In this ``reconstructed" regime, large commensurate AB and BA domains that form within the narrow moir\'e network, increasingly push onto one another in the attempt to increase their respective size, thus squeezing  the discommensurations -- genuine AB/BA dislocations that separate them -- narrower and narrower.
Now while the interlayer energy gain $\propto L dw/d\theta$ is regular as $w \to 0$, the deformation energy within the network increases causing $w$ to decreases,  and the intralayer cost $\sim \mu(a/w)^2wL \propto w^{-1}$, diverges.
Thus the network line width $w$ cannot indefinitely contract, especially  in 2D materials where the intralayer bonds are much stronger than the weak interlayer interactions. Ultimately, a balance between interlayer gain and intralayer cost is established at small $\theta$, where $\theta_\mathrm{AA}$, $w$ and 
$\varepsilon_\mathrm{s} \sim a/8w$
saturate at widths generally much larger than the lattice spacing $a$.
(In the simulated graphene bilayer, $w \approx 3.3$~nm $> 10 a$, which agrees with the experimental observations \cite{Kazmierczak.NatMater.2021}).
\\

Above $\theta_f$, these parameters are $\theta$-dependent. Simulations indeed show that for $\theta \gtrsim \theta_f/2$,
\begin{equation}
    \varepsilon(\theta) \sim \theta_\mathrm{AA}(\theta) = \frac{\theta_f}{\theta}
\end{equation}
reflecting the redistribution of global twist of the moir\'e cell from AB/BA domains, that become virtually untwisted, to a large twirling of the AA nodes (Fig.~\ref{fig:scaling}).

Assuming again a scaling $\varepsilon_\mathrm{s,crit} \sim (D/\mu L a)^{1/2}$, the unbuckled state is predicted to become unstable against a generally antiferro buckling below a sharp critical twist angle
\begin{equation}
    \theta_c \simeq \left( \frac{\mu a^2}{D} \theta_f^2  \right)^{1/3} \sim 
    \left( \frac{\Delta E}{D} \right)^{1/3}
\end{equation}
This is our central result, showing that  the critical twist $\theta_c$ where buckling universally sets in is determined by $\theta_f$, the twist angle below which the interplay between interlayer energy gain and the intralayer cost turns nonlinear resulting in a ``reconstructed" geometry with expanding AB/BA domains, inside an increasingly compressed moir\'e network.
Buckling is clearly prominent for small bending stiffness and large $\theta_f$.
The right side equality holds  under the crude approximation $\theta_f \simeq \theta_0$, where
$\theta_0 = \lim_{\theta \to 0} \theta_\mathrm{AA} \sim \sqrt{\Delta E/\mu a^2}$,
and $\Delta E=E_\mathrm{AA}-E_\mathrm{AB}$ is the sliding energy barrier per atom. 
For three homo-bilayers, $\theta_f$ and $\theta_0$ extracted from realistic simulations are shown in Table~\ref{tab:table1}.
Eq.~(4) predicts that twisted hBN and $\mathrm{MoS_2}$ homo-bilayers should possess larger and slightly smaller $\theta_c$ respectively compared to graphene, which agrees with simulations (Table \ref{tab:table1}).
The onset of homo-bilayer buckling is a sharp mirror-symmetry breaking phase transition.
The antiferro $(2\times 1)$ buckling periodicity attained by all homo-bilayers reflects the lower bending energy cost relative to the ferro $(1\times 1)$, typical of a shear-perturbed membrane \cite{Wang.prb.2023}.
As a result, the energy of $(1\times 1)$ buckled structure is higher than that of the $(2\times 1)$ structure (Fig.~\ref{fig:energy}).
The prevalence of antiferro structures over ferro ones has also been observed in the monolayer membranes containing dilational impurities or grain boundaries \cite{Carraro.pre.1993,Plummer.prm.2022}.
\begin{figure}[ht!]
\centering
\includegraphics[width=0.8\linewidth]{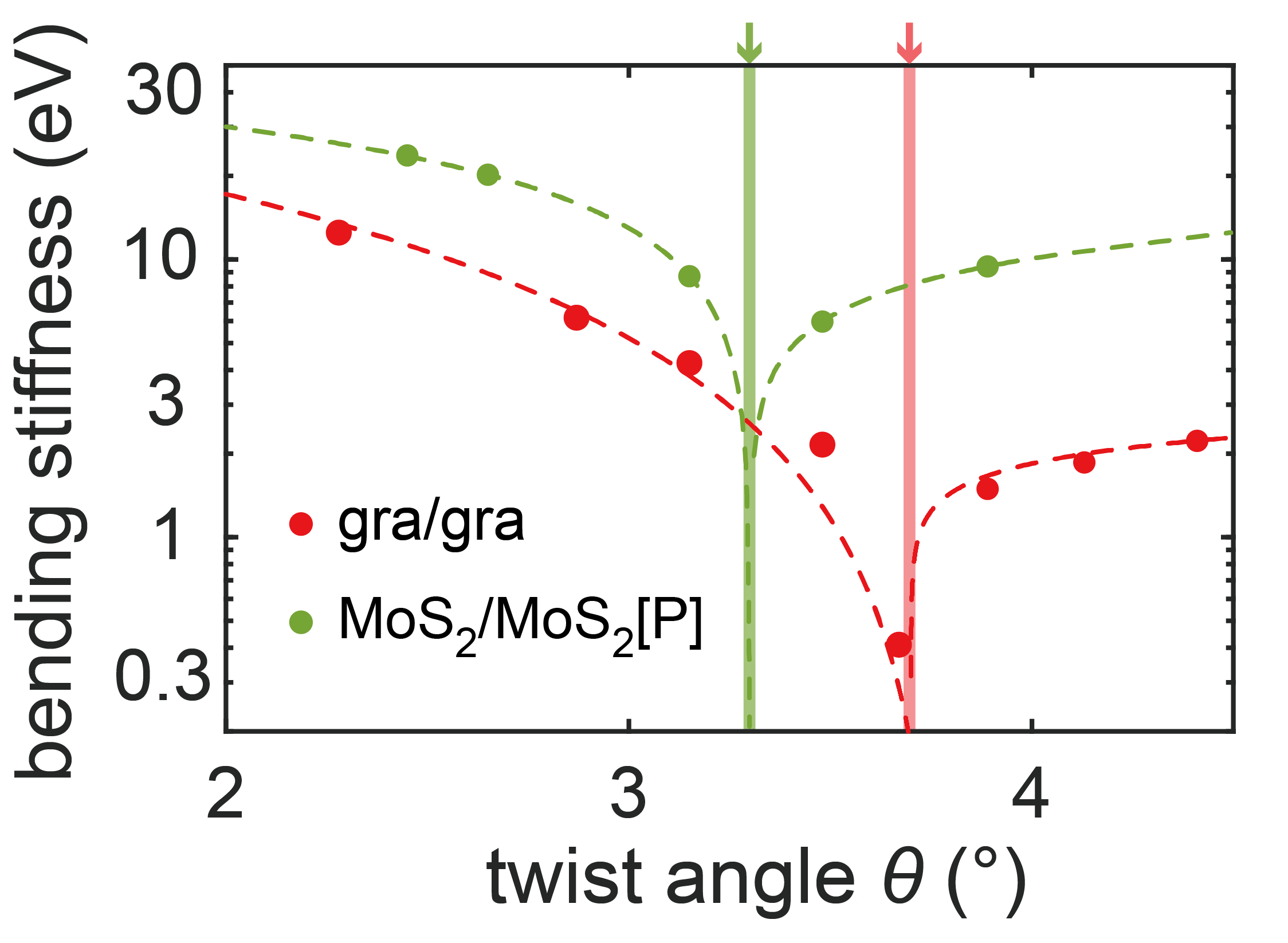}
\caption{Buckling-induced bending stiffness anomaly of two representative homo-bilayers at $\theta$ around $\theta_c$: gra/gra at $3.7^{\circ}$ and P-stacked $\mathrm{MoS_2/MoS_2}$ at $3.3^{\circ}$.
Dashed lines are power-law fits, $D_2 \propto |\theta - \theta_c|^\gamma$. The left and right exponents are respectively $\gamma=$ 1.44 and 0.22 for graphene, and $\gamma=$ 0.57 and 0.38 for $\mathrm{MoS_2}$. 
}
\label{fig:bending}
\end{figure}

In hetero-bilayers, predicting the buckling pattern and magnitude is more complex owing to the superposition of compressive moir\'e stress at all $\theta$ with shear stress growing with twist angle $\theta$.
Lacking a neutral symmetry plane, $(1\times 1)$ buckling will generally occur as a smooth crossover upon decreasing $\theta$. The buckling is preferentially one-sided, upwards or downwards, as we see e.g., in our simulated $\mathrm{WS_2/WSe_2}$ bilayers -- a large $\delta$ case (Fig.~\ref{fig:master}C).
When $\delta$ is sufficiently small however, a $(2\times 1)$ buckled phase may intrude in a narrow $\theta$  range, interrupting the generalized $(1\times 1)$ state (Fig.~\ref{fig:master}B) {an event also encountered in earlier work \cite{Leven.jctc.2016,Mandelli.acsnano.2019,Ouyang.prl.2021}}.
This intrusion occurs in graphene/hBN between $\theta_1 \approx 1.2^{\circ}$, where for increasing $\theta$ the magnitudes of shear and compressive stress become comparable (Fig.\ref{fig:stress}D), and $\theta_2 \approx 2^{\circ}$ where $w$ drops, and the shear perturbing magnitude again dies out. 
Despite the Ising character of the $(2\times 1)$ order parameter, these two transitions are first order as a function of $\theta$, with strain probably entering as a secondary order parameter.
When ideally going from this small-$\delta$ marginal hetero-bilayer to a homo-bilayer where $\delta=0$, $\theta_1 \to 0$ and $\theta_2 \to \theta_c$, as in Fig.~\ref{fig:master}A. At the same time, both transitions revert from first to second order. Table~\ref{tab:table1} {summarizes the properties of the spontaneously buckled $T$=0 structure of each bilayer examined.}

\begin{table*}[ht!]
\centering
\caption{\label{tab:table1} Unbuckling twist angle, effective corrugation $h_\mathrm{eff}$ and other properties of chosen bilayers at some $[\theta]$ values.}
\begin{tabular}{lrrrrr}
Bilayer & gra/gra & hBN/hBN & $\mathrm{MoS_2/MoS_2}$ & gra/hBN & $\mathrm{WS_2/WSe_2}$ \\
\midrule
$\theta_c$ or $\theta_u$ (deg) & 3.7 & 6.0 [P], 5.5 [AP] & 3.3 [P], 3.1 [AP] & 5.9 & 5.2 [P] \\
$h_\mathrm{eff}$  ($\mathrm{\AA}$) & 9.3 [$1.08^{\circ}$, $2\times1$] & 8.0 [$1.47^{\circ}$ AP, $2\times1$] & 10.6 [$1.61^{\circ}$ P, $2\times1$] & 8.5 [$0^{\circ}$, $1\times1$] & 1.8 [$0^{\circ}$, $1\times1$] \\
 & -0.11 [$3.89^{\circ}$, U] & -0.11 [$6.01^{\circ}$ AP, U] & 9.9 [$1.61^{\circ}$ AP, $2\times1$] & 4.1 [$1.46^{\circ}$, $2\times1$] & -0.096 [$7.24^{\circ}$, U] \\ 
 & & & -0.30 [$3.61^{\circ}$ AP, U] & -0.090 [$7.99^{\circ}$, U] & \\
$T_c$ (K) & 540 [$1.08^{\circ}$] & 260 [$1.47^{\circ}$ AP] & $> 600$ [$1.61^{\circ}$ AP] & 160 [$1.46^{\circ}$] & {/} \\
$\sigma_c$ (GPa) & 1.6 [$1.08^{\circ}$] & 0.95 [$1.47^{\circ}$ AP] & 0.46 [$1.61^{\circ}$ AP] & {/} &{/}\\
$\theta_0$ (deg) & 2.3 & 3.1 [P], 2.6 [AP] & 4.6 [P], 4.2 [AP] & & \\
$\theta_f$ (deg) & 1.4 & 2.5 [P], 1.9 [AP] & 5.3 [P], 4.5 [AP] & & \\
\bottomrule
\end{tabular}\\
{P, AP indicate  parallel or anti-parallel stacking; U indicates unbuckled.}
\end{table*}

\section*{Properties of buckled bilayers}

The universal bilayer buckling scenario just presented for a generality of 2D bilayers applies at zero temperature and in ideally perturbation free conditions.
With no data as yet to be confronted with, it is of conceptual and practical importance to establish the response and property evolution of the predicted buckled states under external conditions.
We do this as a direct extension, including all concepts and methods, of our recent exploratory work for the graphene/graphene bilayer \cite{Wang.prb.2023}, to which we refer to for all details.
Below is a brief {summary of the} effects (described in Fig.~7-10) caused by external perturbations on the generality of bilayers and their buckling properties.

\begin{figure*}[ht!]
\centering
\includegraphics[width=0.92\linewidth]{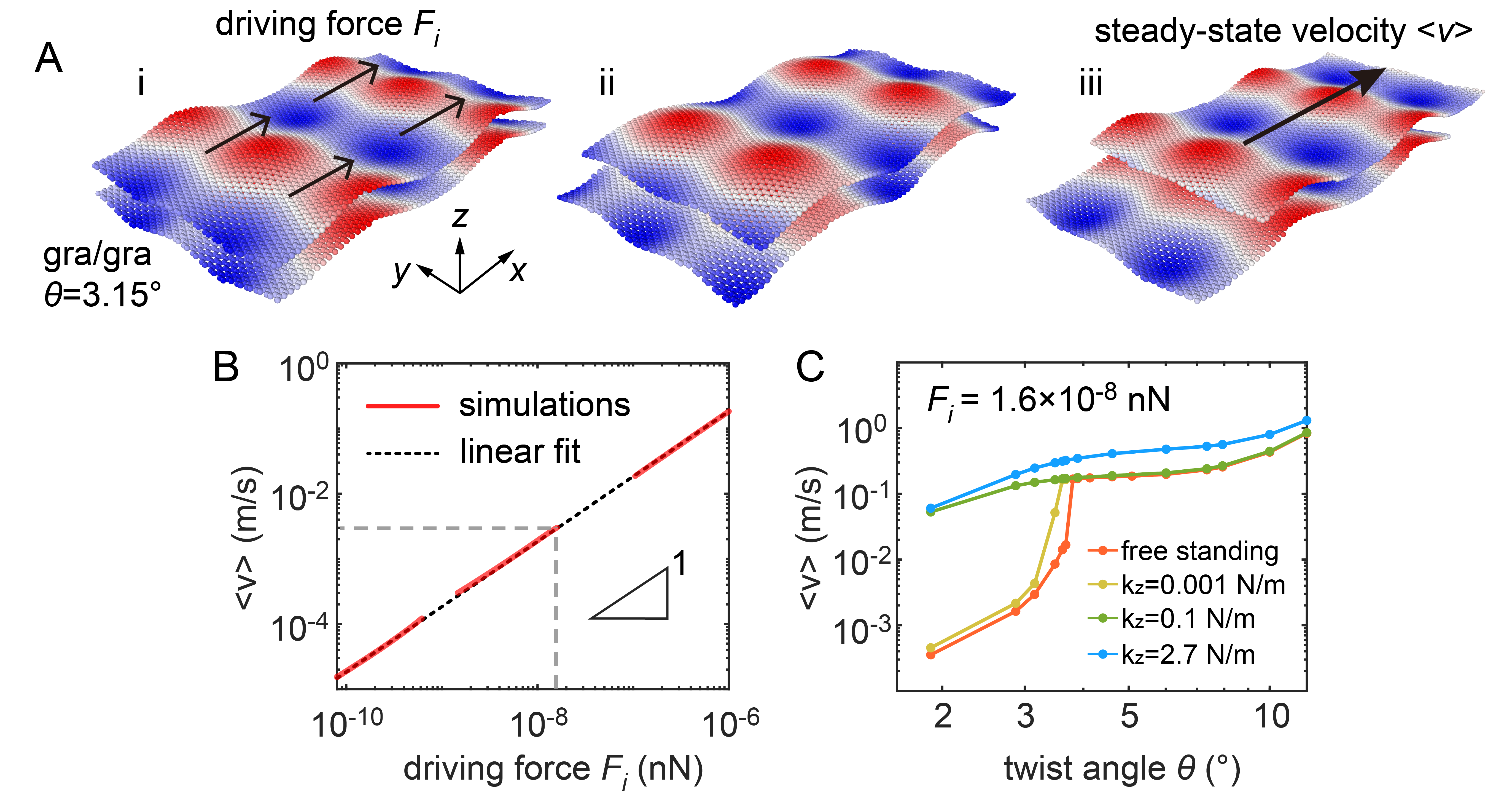}
\caption{A. Simulated sliding of a buckled graphene/graphene bilayer. (i-iii) Cartooned snapshots showing the smooth displacement observed for the upper layer. 
B. Steady-state sliding velocity $\langle v \rangle$ versus driving force $F_i$. Linearity down to the lowest force indicates persistence of structural superlubricity despite the buckling. 
C. Evolution of steady-state velocity at very weak driving force while buckling is being gradually suppressed by ``deposition"  on a substrate of increasing stiffness. 
The substrate and its elastic stiffening is {described} by $z$-springs of stiffness $k_z$ attached to the lower layer.}
\label{fig:friction}
\end{figure*}

\textit{Temperature.}
Thermal fluctuations oppose and reduce buckling -- entropy being generally favored by less distorted states. The temperature classical evolution of buckled bilayers is directly obtained by molecular dynamics simulations, performed for all bilayers studied in the previous section.
All $(2\times 1)$ buckled structures, homo and hetero, exhibit an apparently continuous 
buckling-unbuckling phase transition with estimated critical temperatures $T_c$ given in Table I.
We are not equipped here to investigate the associated critical properties. 
Quantitatively, the $(2\times 1)$ order parameter of the low-$\theta$ buckled states examined persists up to rather high temperatures.
Buckling for example survives up to $ T_c \approx$ 500 K in graphene/graphene at the magic twist angle, 260 K in hBN/hBN at $\theta$ = 1.47 $^{\circ}$ (AP stacking), and above 600 K in $\mathrm{MoS_2/MoS_2}$ at 
$\theta = 1.61 ^{\circ}$ (AP stacking).
Such a thermal resilience, surprising at first sight considering the relatively modest buckling energy gain, see Fig~\ref{fig:energy} E, can be traced back to the $(2\times 1)$ phonon mode, whose frequency $\omega(\theta)$, soft at the buckling transition, stiffens back considerably in the buckled state.
That is exemplified for graphene/graphene in Fig.~S5 and S6 of Ref. \cite{Wang.prb.2023},
where the mode stiffening is in turn connected with the associated dynamic stretching of in-plane bonds, very stiff in the generality of 2D materials.

\begin{figure*}[ht!]
\centering
\includegraphics[width=0.8\linewidth]{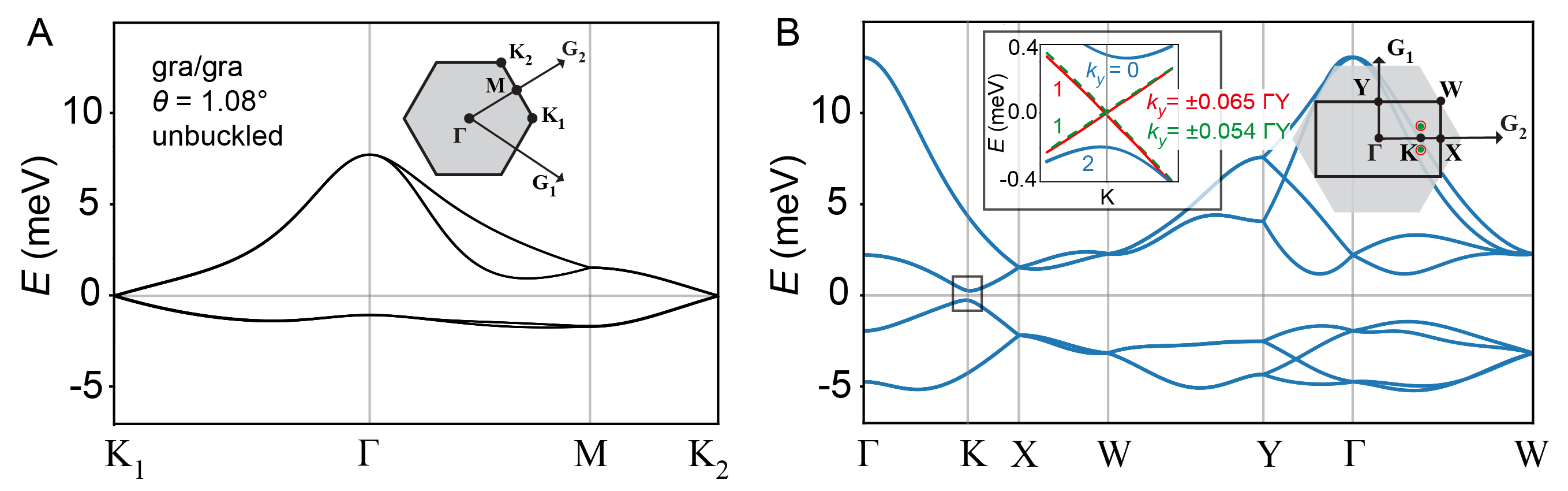}
\caption{Exemplifying  buckling-induced electronic structure changes. Tight binding electronic band structures of a twisted graphene bilayer at magic twist $\theta=1.08^{\circ}$,
{A. supported and unbuckled; B. freestanding and $(2\times 1)$ (``antiferro") buckled (data from Ref.~\cite{Wang.prb.2023}).
Besides the obvious doubling of bands caused by folding and the overall bandwidth increase, note other fine features, such as the apparent gap opening at K. As shown by the inset in (B), there  is in fact no gap. 
The four Dirac points initially folded at K persist, but are split and shifted away from K by a small amount $\pm k_y \neq 0$, suggestive of a pseudomagnetic field effect.}
}
\label{fig:bands}
\end{figure*}

\textit{Bending.}
The bending stiffness $D$, a fundamental mechanical parameter of a membranes, also exhibits a singular softening at the buckling transition, as shown for two homo-bilayers in Fig.~\ref{fig:bending}.
In the simple zigzag model of buckling \cite{Wang.prb.2023}, the bending stiffness is related to 
that, $K$,  of the zigzag ``hinges", in turn connected to the buckling mode frequency $\omega(\theta)$, in the form
\begin{equation}
    D = \frac{D_f}{1+2D_f/\sqrt{3}K}
\end{equation}
where $D_f$ is the bending stiffness of the flat AB/BA region, and the ``hinge stiffness'' $K$ can be expressed as
\begin{equation}
    K \approx \frac{9\rho_\mathrm{2D}a^4}{16\pi^4} \frac{\omega^2}{\theta^4}
\end{equation}
where $\rho_\mathrm{2D}$ and $a$ are the 2D density and lattice constant of the membrane.
The anomalous drop of bending stiffness therefore reflects that of the buckling mode.\\


\begin{figure*}[ht!]
\centering
\includegraphics[width=0.9\linewidth]{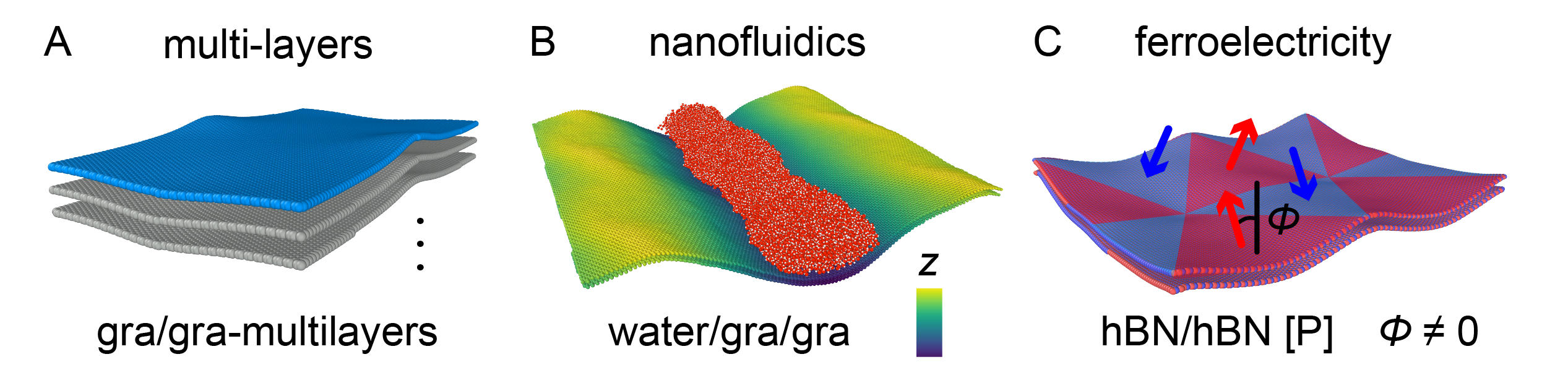}
\caption{ 
Some phenomena  potentially suggested by the buckled state.
{A. Persistence of buckling in thin multilayers \cite{Mandelli.acsnano.2019, Ouyang.prl.2021};}
B. Buckling under liquid deposition (cartoon of water on magic-angle graphene/graphene) suggests potential nanofluidics applications;  
C. Dipole canting in buckled P-stacked hBN/hBN causing weak in-plane antiferroelectricity (sketch).
}
\label{fig:outlook}
\end{figure*}

\textit{External tensile stress.} 
At zero external stress, buckling causes a contraction of the in-plane bilayer area. A tensile applied stress will therefore oppose buckling.
The zero-temperature critical stress $\sigma_c$ that will suppress $(2\times 1)$ buckling as extracted by simulations is reported in Table I (details in SI).
For small twist ($\theta \approx 1^{\circ}$) homo bilayers, the $T$=0 critical stress is of order of one GPa, suggesting that buckling is also robust with respect to residual stresses that should be inevitably present in experimentally suspended geometries.\\

\textit{Sliding lubricity, without and with deposition}
Twisted or otherwise mismatched 2D bilayers are believed to constitute, when ideally flat, infinite, and  defect free, a freely sliding interface, also called ``superlubric'', with zero static friction $F_\mathrm{s}$ \cite{Wang.rmp.2024,Wang.jmps.2023}.
Once going from the flat state to the spontaneously buckled state, will superlubricity survive? Beyond that, how will the interface sliding  mobility $\mu = \langle v \rangle /F_i$ at mean velocity $\langle v \rangle$ be affected by buckling?
Examining just one case, we approached these questions in the prototypical case of buckled bilayer graphene ($\theta=3.15^{\circ}<\theta_c$) in a non-equilibrium molecular dynamics simulation. 
A uniform $x$-directed driving force $F_i$ starting with extremely weak values was applied to all upper layer atoms (Fig.~\ref{fig:friction}A), keeping  the center of mass of the lower graphene layers fixed. 
The buckled bilayer immediately initiated sliding, no matter how small $F_i$ was -- even with $F_i$ as low as $10^{-10}$~nN (Fig.~\ref{fig:friction}B). We found in addition that the $(2\times 1)$ buckling amplitude remained unchanged during sliding, as shown in Fig.~\ref{fig:friction}A.
Considering that a finite size system will always possess some weak pinning, one can take this facile sliding as evidence that {spontaneously buckled bilayer remains superlubric along $x$ despite the buckling.
In further agreement with that, the mean steady-state sliding velocity $\langle v \rangle$ is also found to remain, as in the unbuckled deposited case \cite{Wang.rmp.2024}, strictly proportional to  force. Moreover, once the force was applied along the orthogonal $y$-direction we again found a similar lack of pinning and free sliding, albeit with a somewhat lower velocity (details in SI). We can therefore reasonably conclude that the buckled bilayer remains ideally superlubric in all sliding directions }.

On the other hand, the mobility $\mu = \langle v \rangle /F_i$, is {heavily} affected by buckling, its sharp drop (Fig.~\ref{fig:friction}C) representing in fact a strong effective signal for buckling.
The large increase of interlayer kinetic friction associated with a buckling-induced  mobility drop is easily understood.  A buckled bilayer has a large out-of-plane instantaneous velocity during sliding, and that is well known to increase frictional dissipation \cite{Ouyang.nanolett.2016,Wang.jmps.2023}.
If on the other hand buckling is artificially suppressed -- for example by deposition on a hard substrate mimicked by $z$-directed lower layer constraints -- mobility is found to rebound back to high.
With $z$-springs of stiffness $k_z=2.7$~N/m, a typical value mimicking a bulk  graphite substrate \cite{Guo.JMPS.2012,Wang.acsami.2019}, the mobility remains high across all twist angles, as shown in Fig.~\ref{fig:friction}C.
For much softer springs though, the buckling persisted (in general with a smaller $\theta_c$) and that immediately reduced mobility.
Extending this line of reasoning, we also found that buckling is not limited to bilayers, and can still exist with reduced magnitude 
{in multilayers \cite{Mandelli.acsnano.2019} (Fig.~\ref{fig:outlook}A) and multi-bilayers \cite{Ouyang.prl.2021},}
owing  to the inherent softness of 2D membranes.

\section*{Outlook}
Twisted bilayers (and multilayers) of 2D materials, very popular for their electronic properties, will also develop very unusual mechanical membrane properties when freestanding.
The  moir\'e pattern represents an omnipresent perturbation extending the well established propensity of all membranes to buckle under extrinsic compressive impurities, \cite{Plummer.pre.2020,Plummer.prm.2022,Hanakata.prl.2022} to  this internal and largely shear perturbation. All bilayers, homo and hetero, are predicted to undergo spontaneous moir\'e-periodic bucklings, with either ``ferro" or ``antiferro" periodic ordering among the AA nodes.
The rich scenario which theory and simulation predicts, summarized by Fig.~\ref{fig:master}, is encouraging for future studies, both theoretical and experimental -- the latter to be achieved in suspended geometries \cite{Butz.nature.2014,Wang.prl.2017,Hou.nc.2021}. 
In addition to the properties predicted above in quantitative detail, additional and diverse phenomena may speculatively be expected to take place in buckled bilayers.

\textit{Electronic properties.}
Raising much excitement in recent years in flat deposited bilayers \cite{Cao.nature.2018-1,Cao.nature.2018-2}, electronic properties should be importantly affected by buckling. 
Firstly, 
bucklings such as $(2\times 1)$ will cause band folding.
A specific example is the doubling of narrow bands in a magic angle twisted graphene bilayer when going from deposited and flat to freestanding and buckled \cite{Wang.prb.2023}, exemplified in Fig.~\ref{fig:bands}. 
{Secondly, and independent of folding, the buckling-induced periodic deformation modifies the band structure. In bilayers containing at least one graphene layer, the Dirac point nature and position is going to be generally affected by strain and curvature in interesting manners.
An illustrative example is offered by the narrow band behaviour in magic angle twisted bilayer graphene (TBG) \cite{Wang.prb.2023}, highlighted in Fig.~\ref{fig:bands}.
When going from flat $(1 \times 1)$ to  buckled $(2 \times 1)$,  the four Dirac points initially folded at 
$\mathrm{K} = (2/3)\mathrm{\Gamma X} = (2\pi/(3a), 0 )$,
are  split and shifted off the high symmetry $\mathrm{\Gamma X}$ line by $(0, \pm k_y )$, with $k_y$ = 0.065 and 0.054 in units of $\mathrm{\Gamma Y} = \pi/(\sqrt{3} a)$.
This buckling-induced  Dirac shifting, suggestive of a pseudomagnetic field effect introduced by curvature \cite{Guinea.prb.2010, Castro-Villareal.2017} represents an interesting future line of investigation.}
Finally, electron or hole doping could be expected to extend and reinforce these effects through the further charge inhomogeneity caused by buckling.
The impact of these buckling-induced changes to the overall electronic 
scenario and transport properties of bilayers represents a interesting open question in the context of ``twistronics".

\textit{Nanofluidics.} The persistence of buckling of some bilayers well above room temperature potentially permits nanofluidic applications \cite{Ma.natmater.2016,Mouterde.nature.2019}, with adsorbed liquids modifying and possibly reinforcing the buckling magnitudes as cartooned in Fig.~\ref{fig:outlook}B.

\textit{Ferroelectricity.} In the context of ferroelectric bilayers, one can anticipate that for parallel-stacked hBN and TMDs \cite{Stern.science.2021,Weston.natnano.2022}, buckling will cause a canting of the polarization vectors by an angle $\Phi$ (Fig.~\ref{fig:outlook}C), giving rise to weak in-plane antiferroelectricity, also changing the sensitivity to external fields.

In closing, we note that related buckling concepts and consequences naturally apply to {thick multilayers, where buckling effects were found in simulations \cite{Ouyang.prl.2021,Mandelli.acsnano.2019}}, as well as to other 1D and 0D systems, such as nanotubes, where buckling has been recently highlighted \cite{Leven.natnano.2016,Cao.jpcl.2024},  and {even} in nanospheres \cite{Nelson.pre.2003,Nelson.pre.2007,Li.prl.2011}.


%
\newpage

\section*{Methods}
\subsection{The monolayer model}
The monolayer model proposed to represent the neutral membrane has a triangular crystalline structure. The interatomic interactions include harmonic bonds, angles, and dihedrals.
The chosen monolayer rigidity:  Young's modulus $Y$, shear modulus $\mu$, and bending stiffness $D$ are close to those of real 2D materials. Specifically, $Y\approx 300$~N/m, $\mu\approx 120$~N/m, and $D\approx2.8$~eV.

\subsection{Simulation of real 2D bilayers}
Homo- and hetero- bilayers with periodic boundary conditions (PBC) along $x$ and $y$ directions are constructed for a discrete set of twist angles $\theta$ and lattice mismatch $\delta$. No additional constraints are imposed along the out-of-plane $z$-direction.
For homo-bilayers, twisted graphene, hBN, and $\mathrm{MoS_2}$ are studied with $\theta$ ranging from $0.66^\circ$ to $30^\circ$. For the last two cases, both parallel (P) and anti-parallel (AP) configurations are considered.
For hetero-bilayers, gra/hBN ($\delta \approx 1.83\%$) and $\mathrm{WS_2/WSe_2}$ ($\delta \approx 5.12\%$) are studied with $\theta$ ranging from 0 to $30^{\circ}$.
The interlayer interaction is described by the registry-dependent Kolmogorov-Crespi potential and registry-dependent interlayer potential \cite{Kolmogorov.prb.2005,Ouyang.nanolett.2018,Ouyang.jctc.2021,Jiang.JPCA.2023}. 
The intralayer interaction for carbon, hBN, and TMDs are described by the second-generation reactive empirical bond order potential, Tersoff, and Stillinger-Weber potentials respectively \cite{Brenner.jpcm.2002,Kinaci.prb.2012,Jiang.amss.2019}.
All simulations are performed with {the} open-source code LAMMPS \cite{Plimpton.jcp.1995,Thompson.compphyscomm.2022}.

For structural optimizations,  the simulation box adaptively changes size, so that the in-plane stress is fixed to zero, $p_{xx}=p_{yy}=0$. 
The FIRE algorithm \cite{Bitzek.prl.2006} is used to minimize energy during structural optimization (together with CG algorithms with several loops to optimize the box size).
Minimization stopped when the largest single atom force $|f_i| < 10^{-6}~\mathrm{eV/\AA}$.

A compression protocol and related techniques are used to extract the bending stiffness of 2D bilayers \cite{Wang.prb.2023}. 
Other simulations, including sliding simulations to extract the static and kinetic friction, quasi-static tensile simulations to predict critical tensile stress $\sigma_c$, and equilibrium dynamics simulations with adsorbed water, are provided in the supplementary information.

\section{Acknowledgments}
Work carried out under ERC ULTRADISS Contract No. 834402, with discussions with E. Meyer and M. Kisiel.
J.W. acknowledges the computing resources support from National Supercomputer Center of Tianjin.

\section{author declaration}
The authors declare no competing interest.


\bibliography{ref.bib}

\end{document}



\title{Supplemental Material \\
Universal Moir\'e Buckling of \jin{Freestanding} 2D Bilayers}

\author{Jin Wang}
\affiliation{International School for Advanced Studies (SISSA), I-34136 Trieste, Italy}
\affiliation{International Centre for Theoretical Physics, I-34151 Trieste, Italy}

\author{Erio Tosatti}
\email{tosatti@sissa.it}
\affiliation{International School for Advanced Studies (SISSA), I-34136 Trieste, Italy}
\affiliation{International Centre for Theoretical Physics, I-34151 Trieste, Italy}


\maketitle

\tableofcontents
\newpage

\section{Free energy of the buckled membrane}

The  Föppl–von Kármán free energy $F$ of a generic $(x, y)$ free-standing membrane
under an external perturbation $U$ \cite{Landau.elasticity}, is
\begin{equation}
\begin{aligned}
    \Pi= F + U &= \frac{1}{2}\iint dx dy \big[ D (\nabla^2 h)^2 + \lambda (\varepsilon_{xx}+\varepsilon_{yy})^2 \\
&+ 2 \mu(\varepsilon_{xx}^2+\varepsilon_{yy}^2+2\varepsilon_{xy}^2) -2d \sigma_{ij} \varepsilon_{ji}  \big ]
\end{aligned}
\end{equation}
where the last term is the coupling to the externally driven stress field $\sigma_{ij}$, $d$ is the effective thickness of the membrane, $h$ is the out-of-plane $z$-deformation, $D$ the bending stiffness, $\lambda$ and $\mu$ are the 2D lam\'e constants, and strain
\begin{equation}
    \varepsilon_{ij} = \frac{1}{2} \left[ \frac{\partial u_i}{\partial x_j} +  \frac{\partial u_j}{\partial x_i} + \frac{\partial h}{\partial x_i} \frac{\partial h}{\partial x_j} \right]
\end{equation}
with $u_x$ and $u_y$ the in-plane elastic deformations.

\subsection{Uniform compression}

We consider firstly the effect of uniform compression applied to a square membrane with side length $L$, as illustrate in main text Fig.~1(a). The displacement fields can be approximated by:
\begin{equation}
\begin{aligned}
    u_x &= \delta_1 (1-x/L) \\
    u_y &= \nu \delta_1 y/L \\
    h &= h_0 \sin(2\pi x/L)
\end{aligned}
\end{equation}
where $\nu$ is the Poisson's ratio, $\delta_1$ is the compressive displacement along $x$, and $h_0$ is the corrugation height.
The external work is
\begin{equation}
    U=-N_{xx} \delta_1 L,
\end{equation}
where $N_{xx}=\sigma_{xx} d$ is the compressive load. One can get the minimum of the total potential energy by solving
$\partial \Pi / \partial \delta_1 = 0$ and $\partial \Pi / \partial h_0= 0$, which leads to two solutions:

For the first one, we have $h_0 = 0$ and $N_{xx}=Y \delta_1/L$, where $Y$ is the (2D) Young's modulus. This result corresponds to the pre-buckling regime where the membrane is still flat. While for the second one, a second order buckling transition occurs at
$\delta_1 = \delta_\mathrm{1crit}=4\pi^2 D/Y L$ to give rise to a finite corrugation
\begin{equation}
    h_0 = \mathrm{Re}\left[ \frac{2(1-\nu^2)}{3\pi^2} \left( \delta_1 L - \frac{4\pi^2 D}{Y} \right) \right]^{1/2} 
\end{equation}
The corresponding optimized free energy is
\begin{equation}
    F_\mathrm{c} = \frac{Y \delta_1^2}{2} - \frac{Y(1-\nu^2)}{3}\left( \delta_1 - \frac{4\pi^2 D}{Y L} \right)^2 \Theta{\left( \delta_1 - \frac{4\pi^2 D}{Y L} \right)}
\end{equation}

\subsection{Uniform shear}

Now let us consider the effect of uniform shearing applied to the same square membrane, as illustrated in main text Fig.~1(b). The displacement fields are:
\begin{equation}
\begin{aligned}
    u_x &= \delta_2 y/2L\\
    u_y &= \delta_2 x/2L\\
    h &= h_0 \sin[2\pi(x-y)/L]
\end{aligned}
\end{equation}
where $\delta_2$ is the shear displacement. The external work can be expressed as
\begin{equation}
    U = - N_{xy} \delta_2 L
\end{equation}
where $N_{xy}=\tau_{xy}d$ is the shear load. The minimum $\Pi$ is obtained by solving $\partial \Pi / \partial \delta_2 = 0$ and $\partial \Pi / \partial h_0= 0$, which leads to two solutions, similar to that in the compression case:

The first pre-buckling solution gives $h_0=0$ and $N_{xy}=\mu \delta_2/L$, where $\mu$ is the (2D) shear modulus. 
The second buckling solution gives
\begin{equation}
    h_0 = \mathrm{Re} \left[ \frac{1-\nu}{6\pi^2} \left( \delta_2 L - \frac{8\pi^2 D}{\mu L} \right) \right]^{1/2}
\end{equation}
The corresponding optimized free energy is
\begin{equation}
    F_\mathrm{s}= \frac{\mu \delta_2^2}{2} - \frac{\mu(1-\nu)}{6} \left( \delta_2 - \frac{8\pi^2D}{\mu L} \right)^2 \Theta \left(\delta_2 - \frac{8\pi^2D}{\mu L} \right)
\end{equation}

\clearpage
\section{Monolayer model}

{\it Set-ups.}
The non-uniform stress in the model is introduced through a moir\'e-like 2D network of intrinsic in-plane ``impurities'', created by modifying the equilibrium length $l_\mathrm{equ}$ of bonds within the network.
To introduce local shear, two adjacent bonds at $120^{\circ}$ are changed from
$a$ to $l_\mathrm{equ}=(1+\varepsilon_\mathrm{s}) a$ and $(1-\varepsilon_\mathrm{s}) a$ respectively.
To introduce compression, bonds are changed to $l_\mathrm{equ}= (1-\varepsilon_\mathrm{c})a$, compensated by a small increase in the rest interior regions to ensure unchanged overall initial size.

\begin{figure*}[ht!]
\centering
\includegraphics[width=0.8\linewidth]{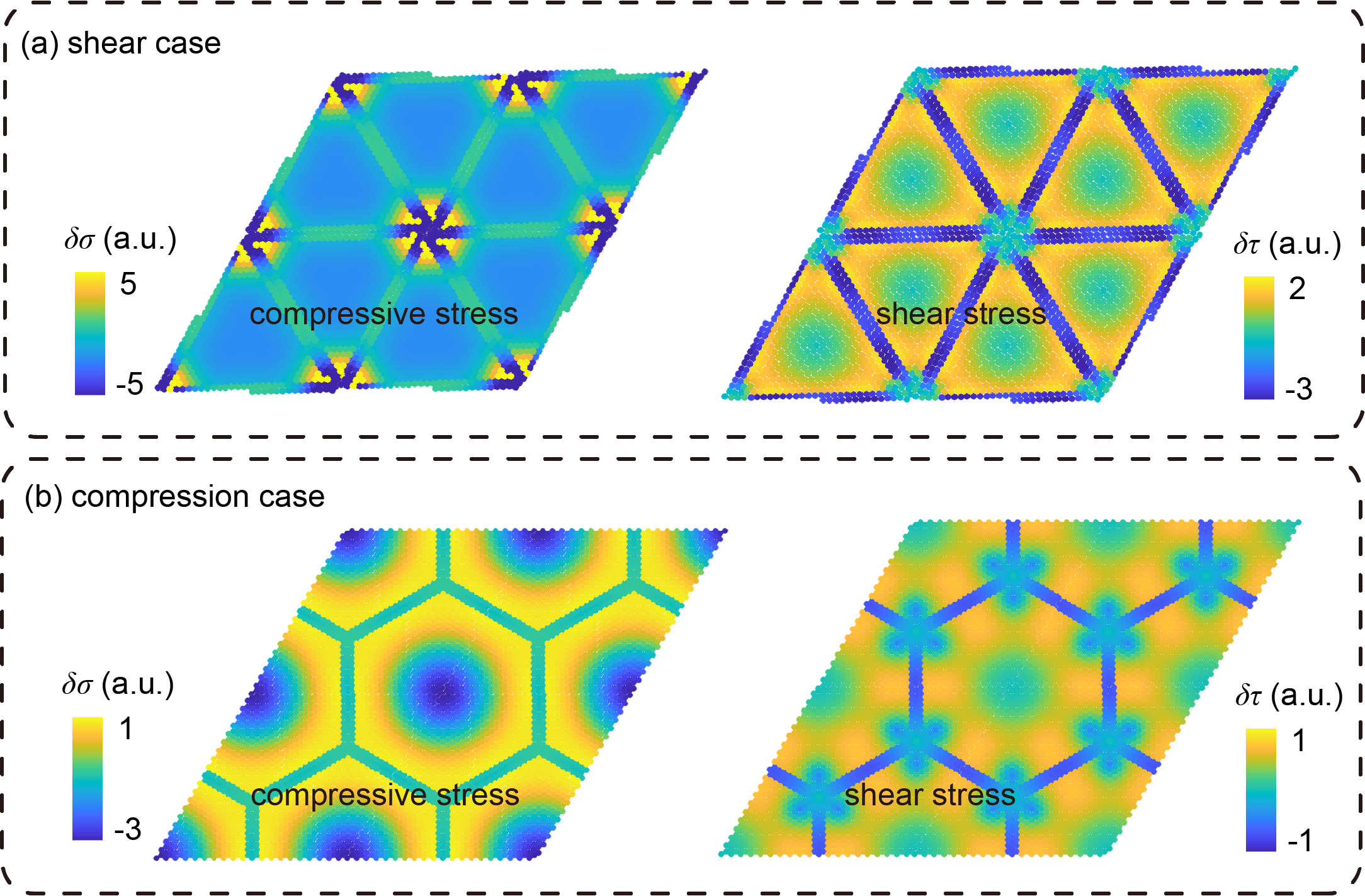}
\caption{Difference in the compressive and shear stress fields for (a) shear and (b) compression cases. Parameter $\delta \sigma$ and $\delta \tau$ are defined in text.}
\label{fig:mono_stress}
\end{figure*}

{\it Stress difference between buckled and flat structure.}
Under the aforementioned set-ups, when $\varepsilon$ is sufficiently large, the optimized structure of the monolayer will buckle. For shear and compressive case, the release of the elastic energy during buckling is proportional to $L^1$ and $L^2$ respectively.
This is shown from the difference in compressive and shear stress ($\delta \sigma$ and $\delta \tau$) before and after buckling:
\begin{equation}
    \begin{aligned}
        \delta \sigma &= (\sigma_\mathrm{buckled} - \sigma_\mathrm{flat}) ~\mathrm{sgn}(\sigma_\mathrm{flat}) \\
        \delta \tau &= (\tau_\mathrm{buckled} - \tau_\mathrm{flat} )~\mathrm{sgn}(\tau_\mathrm{flat})
    \end{aligned}
\end{equation}
where $\sigma = \sigma_1 + \sigma_2$ and $\tau = (\sigma_2 - \sigma_1)/2$, $\sigma_1$ and $\sigma_2$ are two principal stresses ($\sigma_2 > \sigma_1$).
Based on this definition, negative $\delta \sigma$ and $\delta \tau$ represent the release of stresses, which correspond to energy gain.

As shown in Fig.~S1, in shear case, the energy gain mainly comes from the release of the shear stress within narrow networks, whereas for compression case, it comes from the release of compressive stress within large interior regions of the network.

\section{Corrugations in simulated real bilayers}

\begin{figure*}[ht!]
\centering
\includegraphics[width=0.8\linewidth]{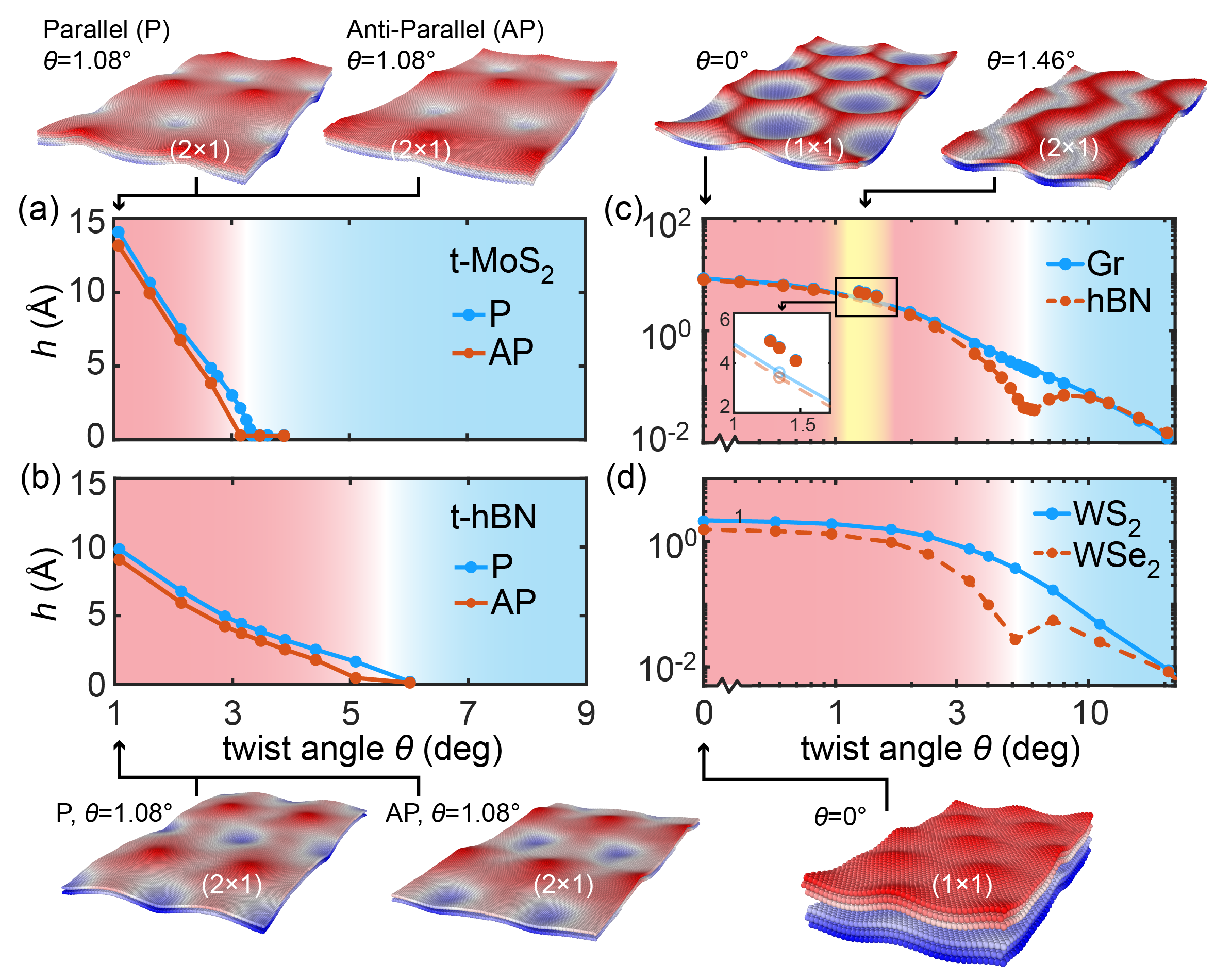}
\caption{Optimized structure and their corrugation amplitude $h=\max{\{|\Delta z|\}}$ for various real bilayers at different twists. 
For homo-bilayers we show results for both parallel (P) and anti-parallel (AP) stacking. The local minima shown in the two heterostructures is due to a change of the sign, corresponding to $\theta_\mathrm{u}$ in Fig.~6 of \jin{the} main text.}
\label{figS:structures}
\end{figure*}

Here in Fig.~\ref{figS:structures} we show the optimized structure and corrugation height $h=\max\{ |\Delta z| \}$ of four bilayers at different twists, including $\mathrm{MoS_2}$ and hBN homo-bilayers, as well as gra/hBN and $\mathrm{WS_2/WSe_2}$ hetero-bilayers.

\begin{figure*}[ht!]
\centering
\includegraphics[width=0.95\linewidth]{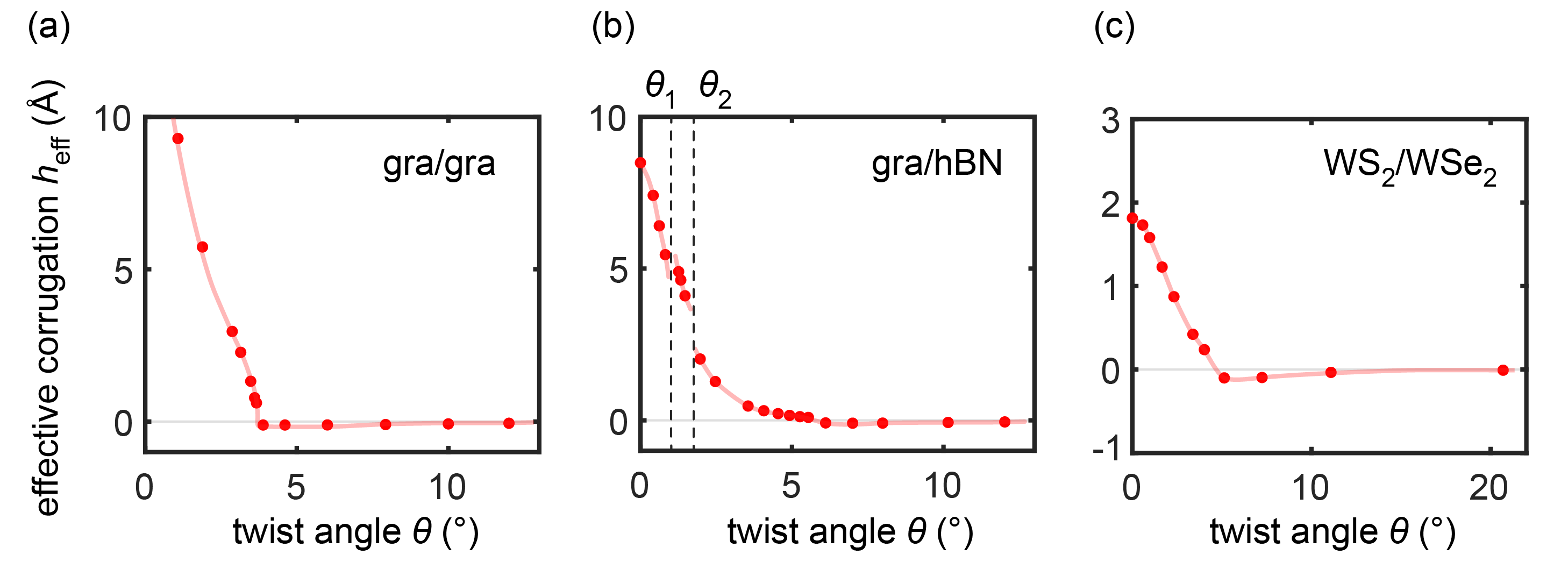}
\caption{
{Evolution of the simulated effective corrugation $h_\mathrm{eff}$ with respect to the twist angle $\theta$. (a) graphene/graphene homo-bilayer; (b) graphene/hBN hetero-bilayer; (c) $\mathrm{WS_2/WSe_2}$ hetero-bilayer. 
Buckling-unbuckling occurs at the twist angle where $h_\mathrm{eff}$ crosses from positive to negative. When the sign change occurs with a finite slope, buckling-unbuckling is a smooth crossover. 
Genuine phase transitions, either from buckling to unbuckling as in gra/gra, or from one buckling symmetry to another, as in gra/hBN, occur when the slope $dh_\mathrm{eff}/d\theta \to \infty$.}
}
\label{figS:effective_h}
\end{figure*}

{As seen in Fig.~\ref{figS:structures}, for heterostructures as twist angle increases, the corrugation height $h$ of one layer gradually decreases, while the other layer exhibits a local minimum. 
This occurs because, as $\theta$ increases, the height $h$ diminishes to zero and then reverses sign, marking a smooth transition from the $(1\times 1)$ buckling to the flat structure. 
To illustrate more clearly this ``reversal'', we extended the definition of buckling corrugation by combining the corrugation heights of both layers and incorporating the direction of the two corrugations, resulting in the following expression for the effective corrugation:}
\begin{equation}
    h_\mathrm{eff} = 2\sqrt{\max{|\Delta \vec{z}_\mathrm{up} \cdot \Delta \vec{z}_\mathrm{low}|}} ~\mathrm{sign}(\left <  \Delta \vec{z}_\mathrm{up} \cdot \Delta \vec{z}_\mathrm{low} \right >)
\end{equation}
{where $\Delta \vec{z} = \vec{z} - \left<\vec{z} \right>$. Based on Eq.~S12, positive values of $h_\mathrm{eff}$ corresponds to buckled states; while negative values correspond to flat, unbuckled states.
The $h_\mathrm{eff} (\theta)$ relationship for three representative 2D bilayers is shown in Fig.~\ref{figS:effective_h}.}

\section{Estimation of $\theta_c$ in homo-bilayers}

\begin{figure*}[ht!]
\centering
\includegraphics[width=1\linewidth]{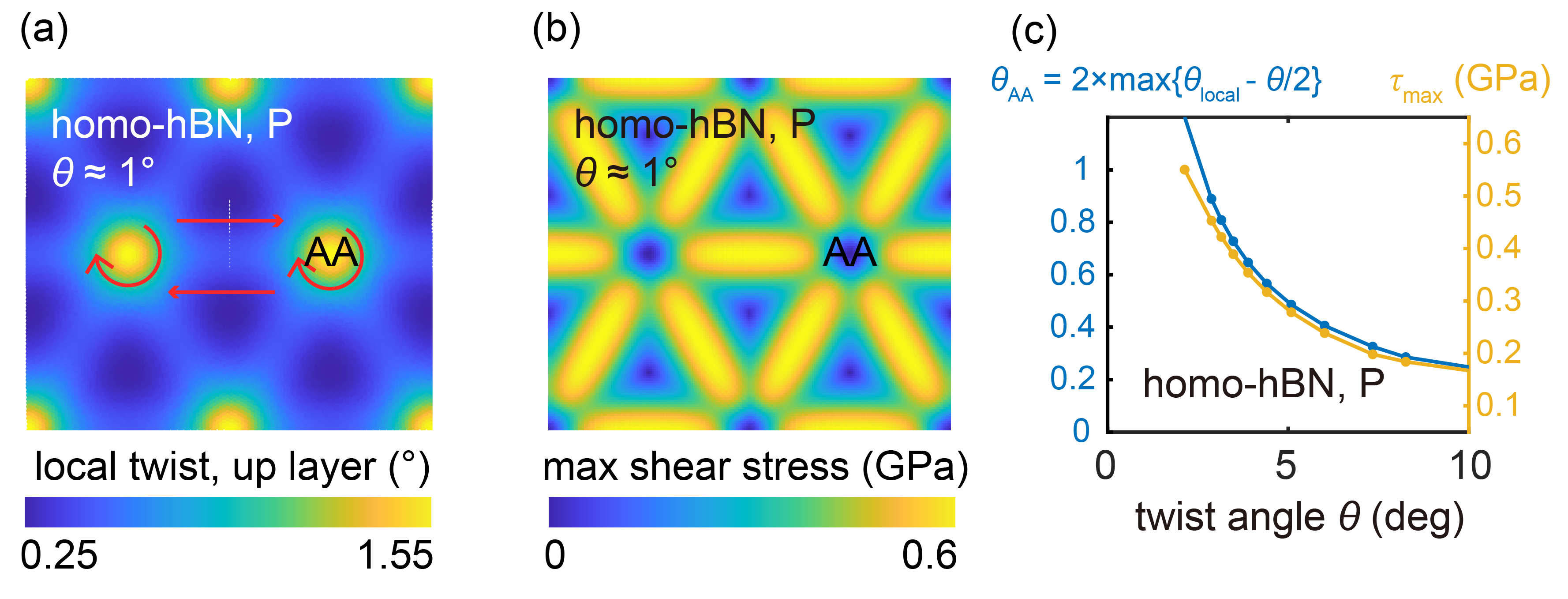}
\caption{The local twist $\theta_\mathrm{local}$ (a) and the maximum shear stress $\tau_\mathrm{max}$ (b) of twisted bilayer hBN at $\theta \approx 1^{\circ}$. (c) Comparison between the AA node twist $\theta_\mathrm{AA}$ and $\tau_\mathrm{max}$ for different $\theta$.}
\label{figS:local_compare}
\end{figure*}

In this section we give derivations for critical buckling twist $\theta_c$.
We begin with a simpler version, with a crude approximation:
\begin{equation}
    \varepsilon_\mathrm{s} (\theta) \sim \theta_\mathrm{AA} =\theta_0 \Theta(\theta_f - \theta)
\end{equation}
where $\theta_\mathrm{AA}$ is the AA node twist and 
$\theta_0 = \lim_{\theta\to 0} \theta_\mathrm{AA}$, which can be expressed as 
\begin{equation}
    \theta_0 = \sqrt{\frac{8\pi \Delta E_\mathrm{DS}}{\mu A_i}}
\end{equation}
Here, $\Delta E_\mathrm{DS}$ is the (per-atom) energy difference between the dislocation region and the AB region, $\mu$ is the shear modulus, $A_i$ is the per-atom area, and $\theta_f$ is the flat-tire crossover and $\theta_f \approx 3^{\circ} $\cite{Wang.rmp.2024}.
We show in Fig.~\ref{figS:local_compare} the comparison of the shear stress in the dislocation region and the local twist at AA nodes at different twist $\theta$ of homo-hBN (P stacking), which verifies the rationality of the first approximation in {Eq.~(S13)}.

Next, by assuming the same scaling we found in monolayer model,
\begin{equation}
    \varepsilon_\mathrm{s, crit} (\theta) \sim \sqrt{\frac{D}{\mu L a}} = \sqrt{\frac{D \theta}{\mu a^2}} 
\end{equation}
where $D$ is the bending stiffness, $L$ and $a$ is the lattice constant of moir\'e lattice and atomic lattice respectively, a critical buckling angle $\theta_c$ can be estimated by equating two formulae, resulting in (for $\theta<\theta_f$):
\begin{equation}
    \theta_c \sim \theta_0^2 \frac{\mu a^2}{D} \sim \frac{8\pi\Delta E_\mathrm{DS}}{D} \sim \frac{\Delta E_\mathrm{AA}}{D}
\end{equation}
This concise equation shows that the buckling occurs more easily when bending stiffness is relatively small or interlayer interaction is strong.\\

Below we give a more accurate derivation. Based on simulation results (Fig.~4 in the main text), we find that the shearing at the dislocation can be approximated by
\begin{equation}
    \varepsilon_\mathrm{s} (\theta) \sim \theta_\mathrm{AA} =\theta_f / \theta
\end{equation}
Thus, the critical angle can be expressed as:
\begin{equation}
    \theta_c \sim \left[ \frac{\mu a^2}{D} \theta_f^2  \right]^{1/3}
\end{equation}
Compared to the previous {Eq.~S16}, this result reveals the effect also {of} the stiffness and $\theta_f$ of the 2D membrane.

\clearpage
\section{Bending stiffness of twisted bilayer $\mathrm{MoS_2}$}

For homo-bilayers, the bending stiffness of the bilayer $D$ has three regimes for different $\theta$, as we discovered in twisted bilayer graphene for the first time \cite{Wang.prb.2023}.
For $\theta > \theta_c$, the bending stiffness of the bilayer equals to twice the bending stiffness of a monolayer ($D=2 D_1$) due to the free interlayer sliding. For $0<\theta < \theta_c$, there is a strengthening effect, causing the overall bending stiffness {to be} significantly greater than $2D_1$.
The most interesting regime is when $\theta \to \theta_c$, the bending stiffness of the bilayer has a collapse $\lim_{\theta \to \theta_c} D = 0$ -- a counter-intuitive outcome that the bending stiffness of the bilayer can be smaller than that of a monolayer.
This can be understood in a zigzag model \cite{Wang.prb.2023}, which
{sketches the up-down morphology of the $(2\times 1)$ buckled bilayer as accordeon-like, with flat areas separated by up-down ``hinges" where the bending concentrates (see, e.g., Fig.~S4 in Ref. \cite{Wang.prb.2023}).
In this model, the collapse of the bilayer bending stiffness is due to the weakening of the ``hinge'' stiffness $K$ when $\theta \to \theta_c$ in the $(2\times1)$ buckled structures. As a result, the  intrinsic buckling mode frequency $\omega$ of the structure also drops to zero when $\theta \to \theta_c$ \cite{Wang.prb.2023}. 
This interesting regime, resulting from the interplay between the interlayer and intralayer interactions, naturally does not exist in a 2D monolayer.}

For homo $\mathrm{MoS_2}$ bilayers in our simulations, the critical twist is $\theta_c \approx 3.3^{\circ}$. And again we observe the bending stiffness collapse when $\theta \to \theta_c$. Specific values are listed in the table below. The simulation result of the monolayer agrees with experimental observations \cite{Yu.AdvMat.2021}.

\begin{table*}[ht!]
\caption{\label{tab:table1} Bending stiffness of monolayer and twisted bilayer $\mathrm{MoS_2}$.}
\begin{ruledtabular}
\begin{tabular}{ccccccccc}
System & monolayer & $2.45^{\circ}$ & $2.65^{\circ}$ & $3.15^{\circ}$ & $3.48^{\circ}$ & $3.89^{\circ}$ & $9.43^{\circ}$ \\
\hline
$D$ (eV) & $\approx 10.0$ & 23.7 & 20.2 & 8.71 & 5.98 & 9.44 & 18.03 \\
\end{tabular}
\end{ruledtabular}
\raggedright
\end{table*}

\clearpage
\section{Unbuckling critical temperature and stress}

{The $T=0$ buckling magnitude, represented for example by $h_\mathrm{eff} > 0$, will attenuate in magnitude and eventually disappear at high temperatures. It may also change and disappear under  suitable large mechanical perturbations. In this section, we estimate the buckling-unbuckling transition temperature $T_c$ and external stress $\sigma_c$.}

\begin{figure*}[ht!]
\centering
\includegraphics[width=1\linewidth]{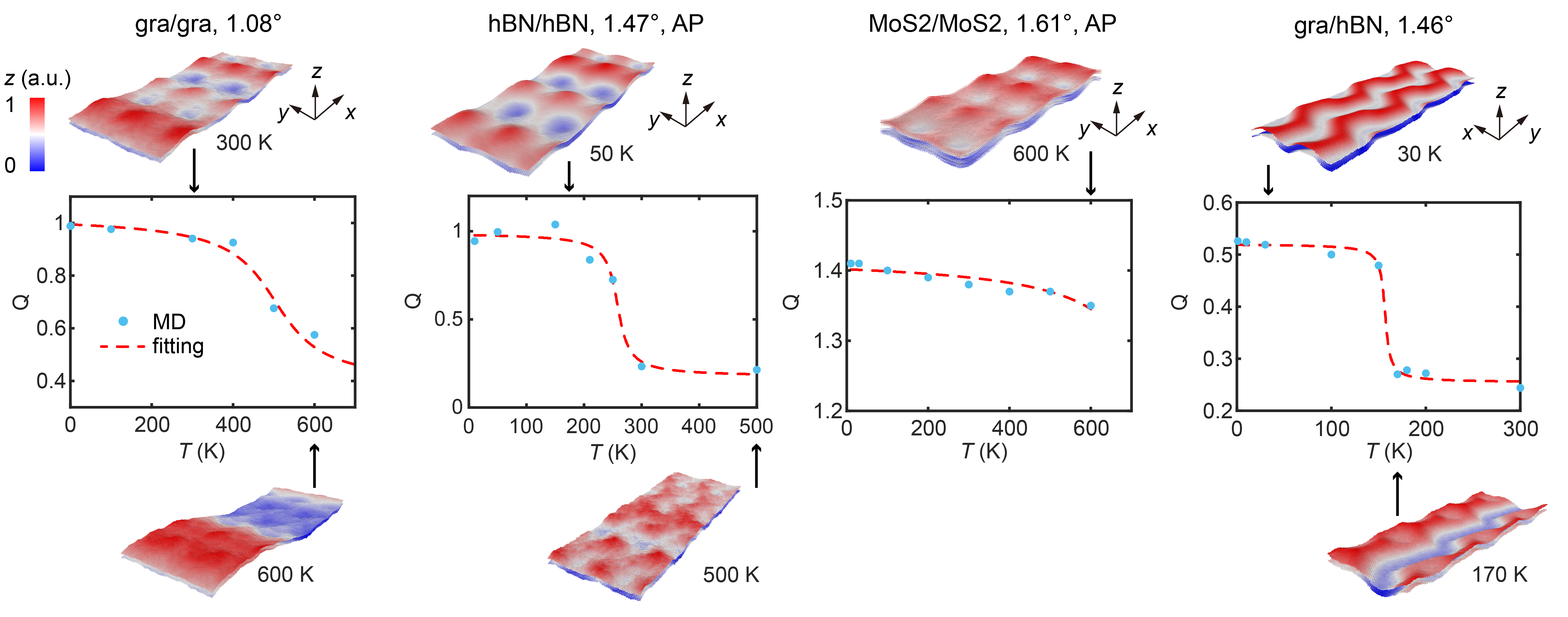}
\caption{$(2\times 1)$ order parameter $Q$ variation as a function of temperature for homo-bilayers. The bilayer changes from buckled to flat when the $T$ is greater than a critical value $T_c$ \cite{Wang.prb.2023}.
{A schematic of the structure at different temperatures (from a single snapshot) is also presented.}
}
\label{figS:critcal_temperature}
\end{figure*}

\jin{As discussed in the main text, we focus on the $(2\times 1)$ buckled structure. Based on its up-down zigzag morphology, the dimensionless buckling order parameter is defined as \cite{Wang.prb.2023}:}
\begin{equation}
    Q = \frac{a}{N_x N_y A} \left< \left| \sum_{n=1}^{N_\mathrm{tot}} z_n \exp(-\frac{2\pi i}{l_x} x_n) \right| \right >
\end{equation}
{where $a$ is the atomic lattice constant, $l_x$ is the spatial distortion periodicity 
along direction $x$, $x_n$ and $z_n$ are coordinates of the $n$-th atom ($n=$1, 2, ..., $N_\mathrm{tot}$), $A$ is the buckled unit cell area, and $N_x$, $N_y$ are the number of cell replicas along $x$ and $y$.
The $(2\times 1)$ buckled structure corresponds to a finite $Q$, while  in the flat (unbuckled) structure  $Q=0$. As defined here, this holds strictly at $T=0$.}

{{\it Temperature.} At finite temperatures, $Q$ is finite in all cases including the flat structures, owing to thermal fluctuations at wavelength $l_x$. Corrections can be made to subtract  off these effects, thus retaining  only the proper buckling order parameter \cite{Wang.prb.2023}. 
An approximate temperature  $T_c$ for buckling's demise can still be extracted from $Q(T)$ by looking for, as shown in Fig.~\ref{figS:critcal_temperature} a significant drop in $Q$.
Ignoring here all proper phase transition issues, a crude value of $T_c$ is obtained by fitting the simulated $Q(T)$  to a shifted and smoothed Heaviside step function (Sigmoid function) $Q(T) = c_1 \{ 1+\exp[(T-T_c)/c_2] \}^{-1} +c_3$. 
Results show that  buckled structures, especially those at small twist angles, have a remarkably large $T_c$, approaching and often exceeding room temperature. Such a large thermal stability highlights the  high possibility for buckling be discovered experimentally at finite temperature, once proper freestanding bilayers were created.
}

\begin{figure*}[ht!]
\centering
\includegraphics[width=0.8\linewidth]{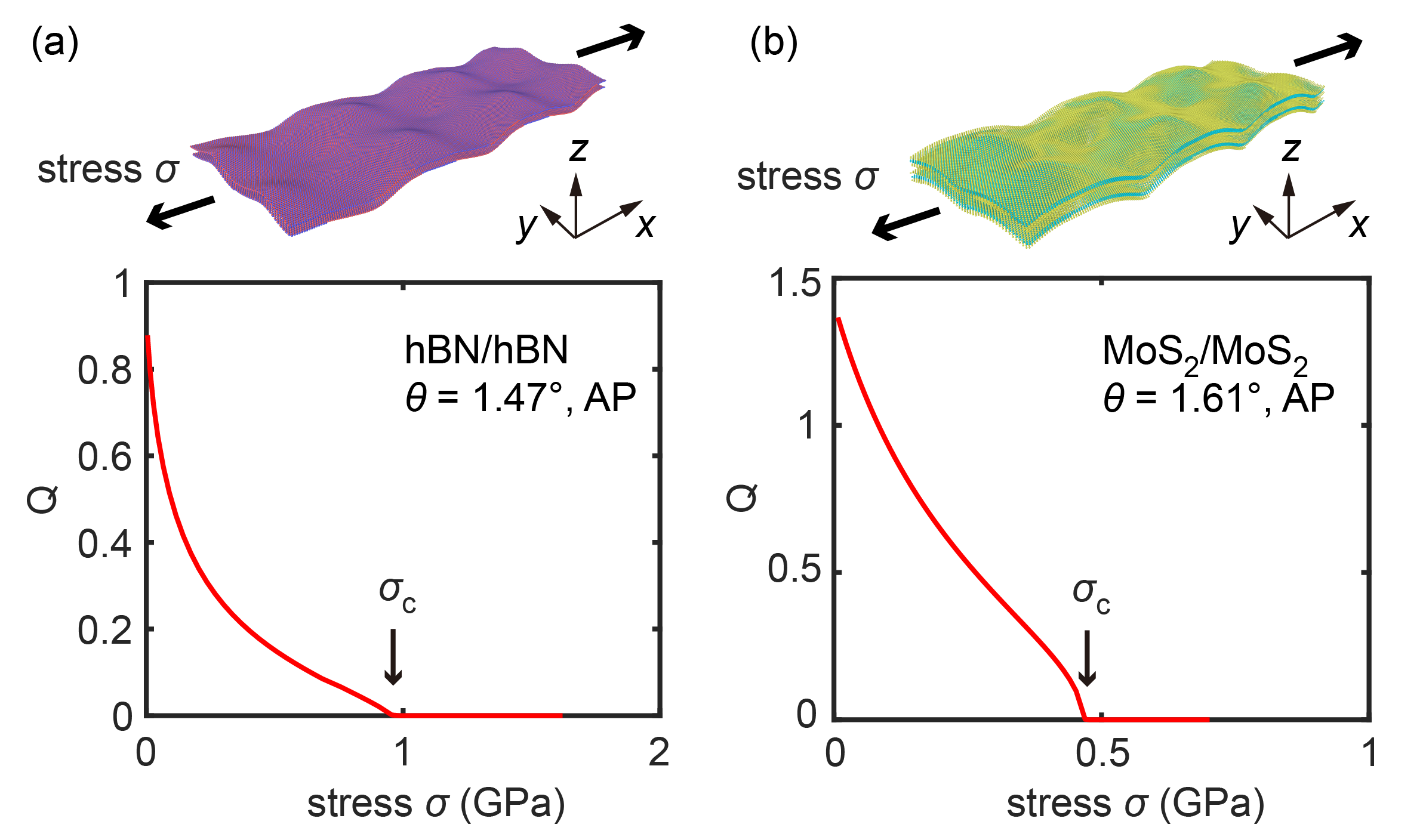}
\caption{Zero-temperature $(2\times 1)$ order parameter $Q$ variation as a function of the external tensile stress $\sigma$ for homo-bilayers. The bilayer changes from buckled to flat when the stress is greater than a critical value $\sigma_c$ \cite{Wang.prb.2023}.
}
\label{figS:critcal_sigma}
\end{figure*}

{{\it Stress.} At zero external stress, buckling causes a contraction of the in-plane bilayer area. A tensile applied stress (along the buckled $x$-direction) will therefore oppose buckling.
The zero-temperature critical stress that will suppress $(2\times 1)$ buckling is studied by means of the elongation simulation.
During the simulation, the entire $(2\times 1)$ buckling structure is gradually stretched along the $x$-direction. The relationship between the $T=0$ order parameter $Q$ and the tensile stress is shown in Fig.~\ref{figS:critcal_sigma}. 
The critical stress $\sigma_c$ is then defined as the point where $Q$ turns from finite  to zero.
For the homo- hBN/hBN and $\mathrm{MoS_2/MoS_2}$ systems shown in Fig.~\ref{figS:critcal_sigma}, as well as the gra/gra system shown in Ref. \cite{Wang.prb.2023}, the critical stress is  quite substantial at small twists, reaching $\sim$GPa. 
This again underlines the high stability of buckling. Of course, if the external stress were compressive instead of tensile, the buckling amplitude would further increase.}

\section{Buckling-related phenomena -- realistic simulations and outlooks}
\textit{Interlayer shearing.}
{We exemplify interlayer shearing by simulations in buckled bilayer graphene. The position of the center of mass of the bottom graphene layer is fixed, and a uniform $x$-directed driving force $F_i$ is applied to each atom of the upper graphene layer. A Langevin thermostat (with isotropic damping coefficient $1$~ps) is applied to the bottom layer to dissipate the frictional energy as needed \cite{Wang.rmp.2024}. 
The simulation timestep is $dt=2$~fs, and the range of $F_i$ used in simulations is from $10^{-10}$~eV/\AA~to  $10^{-6}$~eV/\AA.}\\

\begin{figure*}[ht!]
\centering
\includegraphics[width=0.8\linewidth]{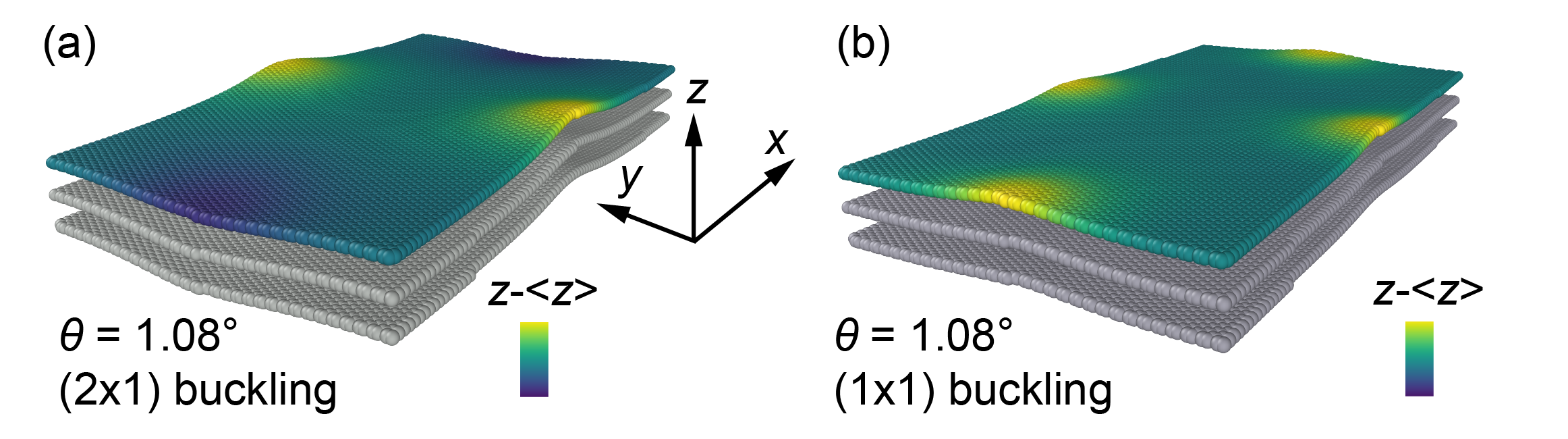}
\caption{Optimized structures for twisted trilayer graphene. (a) $(2\times 1)$ configuration. (b) $(1\times 1)$ configuration.
}
\label{figS:multilayer}
\end{figure*}

{{\it y-shearing.}
For the same $(2\times 1)$ buckled graphene bilayer ($\theta=3.15^{\circ}$), the $y$-directed  frictional driving was also tested (see, e.g., Fig.~\ref{figS:multilayer} for the direction of $x$ and $y$). A small uniform driving force, $F_i=1.6\times10^{-8}$ nN, is applied to all slider atoms along $y$, under similar conditions to the previous $x$-sliding.
Once again, the slider quickly reached a steady state velocity, with an estimated mobility about 60$\%$  smaller than along $x$.
This reduced mobility is rationalised noting that the surfing direction of the moir\'e network is almost perpendicular to the atomic sliding direction \cite{Wang.jmps.2023}. 
When the atoms move along $y$, the moir\'e now surfs along $x$,  causing greater out-of-plane deformation and leading to increased dissipation and smaller mobility. These results along the two main directions are taken to imply that the superlubricity of the system is independent of the sliding direction, although the specific kinetic friction will vary depending on the sliding direction.}\\

\textit{Multi-layer buckling.}
Based on the discussion and understanding in the main text, moir\'e buckling -- although an interfacial feature -- is not limited to bilayers. Theoretically, as long as the twist angle is small enough (resulting in a large moiré size), even thicker materials can undergo overall buckling.
To preliminarily demonstrate this, we established a 1+2 layer model with $\theta=1.08^{\circ}$, where the substrate consists of two layers in AB stacking. Using the same structural optimization method as in the main text, the results showed that the buckled structure can also stably exist in trilayers.
Here in Fig.~\ref{figS:multilayer} we show the optimized structure for $(2\times 1)$ and $(1\times 1)$ buckling configurations.
{Existing studies show that the buckling corrugation can ``penetrate'' into thicker systems \cite{Mandelli.acsnano.2019,Ouyang.prl.2021}.}
\\

\textit{Water nano-river.}
The highly-corrugated structure of the buckling bilayers can naturally serve as channels -- the ``nano-river''. On the other hand, the presence of fluid inside the channel can help stabilize the buckled structure.
To preliminarily demonstrate this, we conducted molecular dynamics simulations based on a $\theta=1.08^{\circ}$ {twisted bilayer graphene (TBG).}
We introduced 6000 water molecules on its upper surface.
{The interaction among water molecules is described by the TIP4P/2005 model \cite{Abascal.jcp.2005}.
A Lorentz-Berthelot mixing rule was adopted with Lennard-Jones parameter $\sigma_\mathrm{C}=3.4872$~\AA~and $\varepsilon_\mathrm{C}=3.0$~meV \cite{Sinclair.advmater.2018} to describe interactions between water and graphene.
The interlayer and intralayer interaction of the bilayer graphene was described by the REBO and ILP force field \cite{Brenner.jpcm.2002,Ouyang.nanolett.2018}, respectively.}
To begin with, we applied a Nose-Hoover thermostat to the entire system for thermal equilibrium ($T=298$~K, $t=0.2$~ns, $dt=1$~fs). Next, we removed the thermostat and allowed the system to evolve freely (NVE) for 2 ns.
{A stable buckled structure persisted during all the simulation, in fact with a strengthened magnitude. In Fig.~10B of main text, we show a snapshot of the structure during the NVE process.}

\clearpage
\bibliography{ref}